\documentclass[useAMS,usenatbib]{mn2e}

\usepackage{deluxetable}
\usepackage{graphicx}
\usepackage{lscape}
\usepackage{changepage}
\usepackage{amsmath}
\title[The Molecular-Hydrogen Kennicutt-Schmidt Law in XUV Disks]{Testing the Molecular-Hydrogen Kennicutt-Schmidt Law in the Low-Density Environments of Extended Ultraviolet Disk Galaxies}

\author[Watson et al.]{Linda~C.~Watson,$^{1,2}$
  Paul~Martini,$^{3,4}$ Ute~Lisenfeld,$^{5}$   
  Torsten~B\"oker$^{6}$ \newauthor and Eva~Schinnerer$^{7}$
  \medskip\\
$^{1}$European Southern Observatory, Alonso de C\'ordova 3107, Vitacura, Casilla 19001, Santiago, Chile; E-mail: lwatson@eso.org\\
$^{2}$Harvard-Smithsonian Center for Astrophysics, 60 Garden Street, Cambridge, MA 02138, USA\\
$^{3}$Department of Astronomy, The Ohio State University, 140 West 18th Avenue, Columbus, OH 43210, USA\\
$^{4}$Center for Cosmology and AstroParticle Physics, The Ohio State University, 191 West Woodruff Avenue, Columbus, OH 43210, USA\\
$^{5}$Departamento de F\'isica Te\'orica y del Cosmos, Universidad de Granada, Spain\\
$^{6}$European Space Agency/STScI, 3700 San Martin Drive, Baltimore, MD 21218, USA\\
$^{7}$Max-Planck-Institut f\"ur Astronomie, K\"onigstuhl 17, D-69117 Heidelberg, Germany}

\begin{document}

\date{Accepted 2015 October 16.  Received 2015 October 15; in original form 2015 January 30}

\maketitle

\begin{abstract}
Studying star formation beyond the optical radius of galaxies allows us to test empirical relations in extreme conditions with low average gas density and low molecular fraction. Previous studies discovered galaxies with extended ultraviolet (XUV) disks, which often contain star forming regions with lower H$\alpha$-to-far-UV (FUV) flux ratios compared to inner disk star forming regions. However, most previous studies lack measurements of molecular gas, which is presumably the component of the interstellar medium out of which stars form. We analyzed published CO measurements and upper limits for fifteen star forming regions in the XUV or outer disk of three nearby spiral galaxies and a new CO upper limit from the IRAM $30 \, {\rm m}$ telescope in one star forming region at $r=3.4 \, r_{25}$ in the XUV disk of NGC~4625. We found that the star forming regions are in general consistent with the same molecular-hydrogen Kennicutt-Schmidt law that applies within the optical radius, independent of whether we used H$\alpha$ or FUV as the star formation rate (SFR) tracer. However, a number of the CO detections are significantly offset towards higher SFR surface density for their molecular hydrogen surface density. Deeper CO data may enable us to use the presence or absence of molecular gas as an evolutionary probe to break the degeneracy between age and stochastic sampling of the initial mass function as the explanation for the low H$\alpha$-to-FUV flux ratios in XUV disks.
\end{abstract}


\section{Introduction}
Studies of star formation in nearby galaxies have divided the relationship between the star formation rate (SFR) surface density ($\Sigma_{\rm SFR}$) and gas surface density ($\Sigma_{\rm gas}$, which includes both atomic and molecular gas) into three regimes, collectively known as the Kennicutt-Schmidt law \citep[$\Sigma_{\rm SFR} \propto \Sigma_{\rm gas}^{N}$;][]{schmidt59,kennicutt98,kennicutt12}. In low density gas ($\Sigma_{\rm gas} \la 10 \, {\rm M}_{\odot} \, {\rm pc^{-2}}$), there is actually little correlation between the total gas and SFR surface density. At somewhat higher densities ($10 \, {\rm M}_{\odot} \, {\rm pc^{-2}} \la \Sigma_{\rm gas} \la 100 \, {\rm M}_{\odot} \, {\rm pc^{-2}}$) where the gas is primarily molecular, the correlation is very good (and $N$ is between 0.8 and 1.5), which confirms that stars form from molecular gas \citep{wong02,bigiel08,rahman12,leroy13}. Finally, merging and starbursting galaxies at low and high redshift have high gas surface densities ($\Sigma_{\rm gas} \ga 100 \, {\rm M}_{\odot} \, {\rm pc^{-2}}$) and are offset to higher SFR surface density at a given gas surface density compared to normal star-forming galaxies \citep{daddi10,genzel10,carilli13}. 

Even in the low-density regime where gas is predominately atomic, recent surveys of molecular gas at $\sim 1 \, {\rm kpc}$ resolution have the sensitivity to probe the strong correlation between the molecular hydrogen surface density ($\Sigma_{\rm H_2}$) and the SFR surface density (the molecular-hydrogen Kennicutt-Schmidt law) down to $\Sigma_{\rm H_2} \sim 3 \, {\rm M}_{\odot} \, {\rm pc^{-2}}$ over a large fraction of the optical disk of nearby spiral and dwarf galaxies \citep{bigiel08,leroy09}. \citet{schruba11} extended the molecular correlation down to $\Sigma_{\rm H_2} \sim 1 \, {\rm M}_{\odot} \, {\rm pc^{-2}}$ by stacking all the CO spectra in $15\arcsec$ rings. This surface brightness limit allowed the authors to study the molecular-hydrogen Kennicutt-Schmidt law out to about the optical radius ($r \sim r_{25}$, where $r_{25}$ is the B-band isophotal radius at $25 \, {\rm mag \, arcsec^{-2}}$).

Studies of star formation beyond the optical radius have primarily concentrated on comparing different star formation tracers. One important discovery was the presence of ultraviolet emission out to four times the optical radius in the disk galaxies M83 and NGC~4625 \citep{thilker05, gildepaz05}. Extended far-UV (FUV) emission indicates the presence of stars with masses greater than $3 \, {\rm M_{\odot}}$ and ages less than $\sim 100 \, {\rm Myr}$. These extended ultraviolet (XUV) disks are found in $4 - 14 \%$ of galaxies out to $z=0.05$ \citep{lemonias11,thilker07}, with a higher fraction in early-type galaxies \citep[39\%;][]{moffett12}.

Radial profiles of the FUV emission in XUV disks generally decline smoothly out to more than twice the optical radius \citep{boissier07,goddard10} and the emission can be primarily diffuse or structured into distinct regions \citep{thilker07}. In H$\alpha$ emission, which traces stars that are more massive than $17 \, {\rm M_{\odot}}$ and younger than $\sim 10 \, {\rm Myr}$, about half of XUV disks show radial profiles that appear truncated near the optical radius and half have radial profiles that are similar to the FUV profile \citep{martin01,boissier07,goddard10}. In general, the XUV disk of a galaxy contains a smaller fraction of the total H$\alpha$ flux than of the total FUV flux because XUV disks have a lower fraction of star forming regions with H$\alpha$ emission compared to the inner disk and the H\,{\small II} regions in XUV disks have lower H$\alpha$-to-FUV flux ratios than the inner disk \citep{goddard10}. 

Because H$\alpha$ and FUV emission generally agree as star formation rate tracers within the optical disk \citep{bigiel08}, a number of authors have investigated the origin of the low H$\alpha$-to-FUV flux ratios in XUV disks. \citet{gogarten09} and \citet{alberts11} performed stellar population synthesis modeling of XUV disk star forming regions with a range of H$\alpha$-to-FUV flux ratios and concluded that age effects alone can explain the low H$\alpha$-to-FUV flux ratios in XUV disks. They measured young ages ($< 10 \, {\rm Myr}$)  for H$\alpha$-bright regions, older ages ($> 16 \, {\rm Myr}$) for FUV-bright, H$\alpha$-faint regions, and a median age for all the star forming regions of $100 \, {\rm Myr}$. \citet{goddard10} studied H\,{\small II} regions in 21 normal and XUV-disk galaxies and concluded that stochastic sampling of the initial mass function (IMF) can explain the H$\alpha$-to-FUV flux ratios in XUV disks because the H$\alpha$ fluxes are consistent with gas ionized by a single O or B0 star \citep[see also][]{gildepaz07}. Such small numbers of high-mass stars indicate that the clusters may not have fully populated the IMF. Similarly, \citet{koda12} concluded that stochastic sampling of the standard Salpeter IMF and aging can explain the number of H$\alpha$- and FUV-bright regions in the XUV disk of M83.

In this paper, we discuss the degeneracy between aging and stochastic sampling of the IMF as the explanation for the low H$\alpha$-to-FUV flux ratios in XUV disks. We use ``stochastic sampling of the IMF'' to mean the incomplete population of the IMF with the rare massive stars that produce ionizing radiation, as is expected in low-mass star forming regions \citep[e.g.,][]{lee09,goddard11,koda12}. Two other potential explanations have been presented in the literature: 1) an IMF with a steep or truncated high-mass end \citep[e.g.,][]{hoversten08, meurer09,lee09}, and 2) leakage of ionizing photons out of the star forming region \citep{elmegreen06, hunter13}. We will not discuss the steep IMF or leakage of ionizing radiation because these explanations have received less support in the literature compared to aging and stochastic sampling of the IMF \citep{goddard10, alberts11,koda12}.

Previous studies of the connection between gas and star formation in XUV disks have primarily focused on atomic gas. \citet{dong08} and \citet{barnes12} concluded that there is a general correspondence between H\,{\small I} and FUV peaks and that the relationship between the SFR surface density and the H\,{\small I} surface density ($\Sigma_{\rm HI}$) is similar to the relationship in the optical disk. A noteworthy exception is the work of \citet{bigiel10}, which showed a correlation between the SFR surface density and H\,{\small I} surface density in the outer disks of 22 nearby galaxies (not necessarily XUV disks), whereas the optical disk shows no correlation.

Most studies of star formation in XUV disks are missing a fundamental component of the star formation process: molecular gas, which is the component of the interstellar medium (ISM) out of which stars actually form \citep[e.g.,][]{fukui10}. There are only four galaxies with molecular gas detections in their outer disks: NGC~4414 \citep{braine04}, NGC~6946 \citep{braine07}, M33 \citep{braine10}, and M63 \citep{dessauges14}. This is likely because the CO surface brightness is low in outer disks and we expect a low covering factor of molecular clouds, given the low covering factor of H\,{\small II} regions and the low molecular cloud counts in the outer disk of the Milky Way \citep{snell12}. \citet{dessauges14} were the first to study the molecular-hydrogen Kennicutt-Schmidt law in an XUV disk. The authors used the Institut de Radioastronomie Millim\'etrique (IRAM) $30 \, {\rm m}$ to carry out twelve pointings on a UV-bright complex at a galactic radius of $r=1.36 \, r_{25}$ in M63. They detected CO(1--0) in two of the twelve pointings and concluded that the molecular gas in those regions has low star formation efficiency (SFE, i.e., low $\Sigma_{\rm SFR}$ at a given $\Sigma_{\rm H_{2}}$) compared to regions in the inner disk.

\begin{figure*}
\begin{center}
\includegraphics[width=0.7\textwidth]{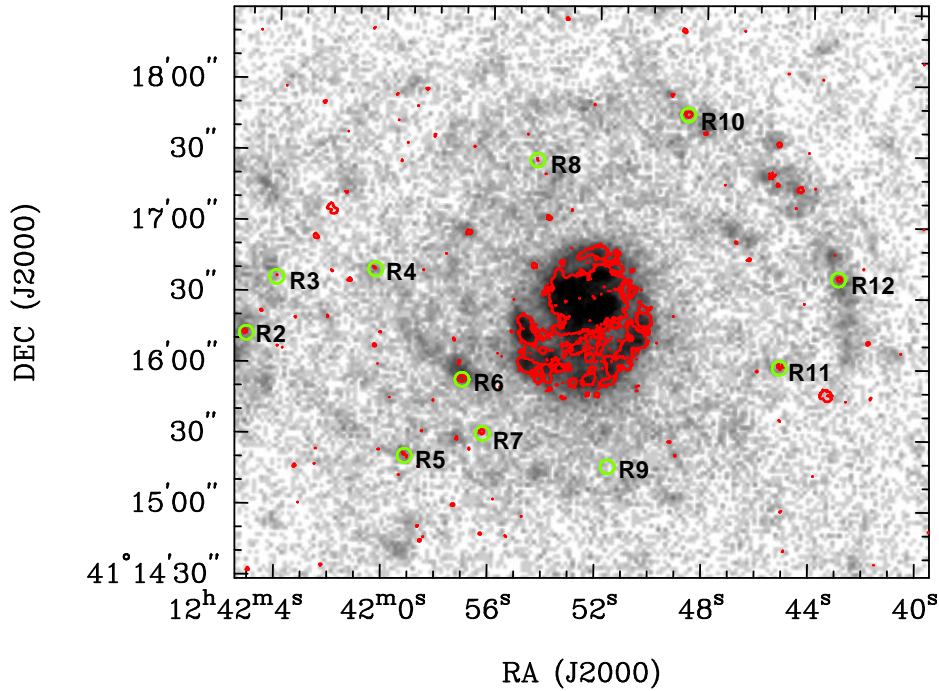}
\caption{Image of NGC~4625, with FUV emission shown as grayscale and
  H$\alpha$ emission shown as a single red contour at four times the
  image rms. The green circles identify the star forming regions
  studied in this work \citep[and originally presented
    in][]{gildepaz07} and have $6\arcsec$ diameter.}
\label{image}
\end{center}
\end{figure*}

We present a study of the connection between molecular gas and star formation in four galaxies, three of which have XUV disks. In NGC~4625, we measured a new, deep CO upper limit for a star forming region at $r = 3.4 \, r_{25}$. In NGC~6946, M33, and NGC~4414, we studied the fifteen regions with $r \geq r_{25}$ and CO measurements or upper limits from \citet{braine07}, \citet{braine10}, and \citet{braine04}, respectively. The goals of the paper are 1) to study star formation in the extreme environment of XUV disks, 2) to determine if the star forming regions are consistent with the same molecular-hydrogen Kennicutt-Schmidt law that applies in the optical disk, and 3) to identify if the presence or absence of molecular gas is consistent with the age of the H\,{\small II} regions as implied by the H$\alpha$-to-FUV flux ratio. The paper is organized as follows. We describe our sample in Section~\ref{sec:sample}. In Section~\ref{sec:data}, we present new CO observations and also measurements derived from published or archival CO, H$\alpha$, FUV, and $24 \, \mu {\rm m}$ data. Section~\ref{sec:results} describes the results of placing XUV disk star forming regions on the molecular-hydrogen Kennicutt-Schmidt law and measuring the ages of the regions. Finally, we discuss our results in Section~\ref{sec:discussion}.

\section{Sample}
\label{sec:sample}
We present a deep upper limit on the molecular gas content in one star forming region in the XUV disk of NGC~4625 ($D = 9.5 \, {\rm Mpc}$; Schruba et al.\ 2011; Figure~\ref{image}). We used the same naming convention as \citet{gildepaz07} and refer to this region as R2. We selected NGC~4625 because it has good ancillary datasets to trace star formation in the XUV disk and we selected R2 for the CO observations because we hypothesized that the relatively high metallicity \citep[$12+{\rm log(O/H)} = 8.5$;][]{gildepaz07} would enhance the likelihood of a CO detection. We also included six regions in the XUV disk of NGC~6946 \citep[$D = 5.9 \, {\rm Mpc}$;][]{schruba11}, six regions in the outer disk of M33 \citep[$D = 0.96 \, {\rm Mpc}$;][]{bonanos06}, and three regions in the XUV disk of NGC~4414 \citep[$D = 17.7 \, {\rm Mpc}$;][]{freedman01}, all of which have deep CO measurements or upper limits \citep{braine04, braine07, braine10}. We refer to the outer disk of NGC~6946 as an XUV disk because the galaxy contains UV emission beyond the optical radius. However, this classification is not as clear as for NGC~4625. Specifically, \citet{holwerda12} classified NGC~6946 as not having an XUV disk. M33 does not have an XUV disk classification in the literature. Outer-disk FUV emission is primarily confined to the northern edge of the galaxy at radii less than about $1.1 \, r_{25}$. We included it in our study because there are so few detections of molecular gas in outer disks. NGC~4414 has an XUV disk that was identified by \citet{thilker07}. Finally, we studied ten additional star forming regions in NGC~4625 (R3 - R12) that have shallow upper limits on the CO intensity from the HERACLES survey \citep{leroy09}. Table~\ref{tab:obs_prop} lists the coordinates and galactocentric radii of the regions we studied.

\section{Data and Measurements}
\label{sec:data}
We used the measurements described in this section for two purposes: to place the XUV disk star forming regions on the molecular-hydrogen Kennicutt-Schmidt law and to measure the ages of the regions. For the Kennicutt-Schmidt law study, we measured molecular hydrogen and SFR surface densities within matched apertures, with the aperture diameter set by the IRAM $30 \, {\rm m}$ beam (${\rm FWHM} = 21\arcsec$ at CO(1--0)). Table~\ref{tab:obs_prop} lists the CO(1--0) and CO(2--1) line intensities, and also the H$\alpha$, FUV, and $24 \, \mu {\rm m}$ fluxes or surface brightness values within a 21$\arcsec$-diameter circular aperture. To estimate the ages of the star forming regions in NGC~4625 and NGC~6946, we measured the H$\alpha$ and FUV fluxes in $6\arcsec$-diameter circular apertures. We present these values in Table~\ref{tab:obs_prop2}.

\subsection{IRAM 30 m Data and Molecular Hydrogen Surface Densities}
We observed R2 in NGC~4625 at the CO(1--0) and CO(2--1) lines at 115 and 230\,GHz on three independent dates (2010 November 19, 2011 March 2/3, and 2011 March 6/7) with the IRAM $30 \, {\rm m}$ telescope on Pico Veleta. The total on-source integration time was 6.2 hours. We used the dual polarization receiver EMIR in combination with the autocorrelators WILMA and VESPA. We carried out our analysis with the VESPA data, but we confirmed that the WILMA data are consistent. The frequency resolution was $0.32 \, {\rm MHz}$ at CO(1--0) (velocity resolution of $0.8 \, {\rm km \, s}^{-1}$) and $1.25 \, {\rm MHz}$ at CO(2--1) (velocity resolution of $1.6 \, {\rm km \, s}^{-1}$). The observations were carried out in wobbler switching mode with a wobbler throw between $220\arcsec$ and $240\arcsec$ in the azimuthal direction. We chose the amplitude of the wobbler throw to avoid star forming regions within the optical disk of NGC~4625 (and the nearby galaxy NGC~4618) in the off-position.

At the beginning of each observing session the center of NGC~4625 was briefly observed in order to check the frequency tuning and the relative calibration. We found good agreement, with differences no greater than $25\%$ in the velocity integrated intensities between the different observations. The line shape and intensities were in good agreement with those presented in \citet{albrecht04}.

Pointing was monitored on nearby quasars every 60 -- 90 minutes. During the observation period, the weather conditions were generally good and the pointing was better than $4\arcsec$. The mean system temperatures were $220 \, {\rm K}$ at $115 \, {\rm GHz}$ and $200 \, {\rm K}$ at $230 \, {\rm GHz}$ on the $T_{\rm A}^*$ scale. At $115 \, {\rm GHz}$ ($230 \, {\rm GHz}$), the IRAM forward efficiency, $F_{\rm eff}$, was 0.95 (0.91), the beam efficiency, $B_{\rm eff}$, was 0.77 (0.58), and the half-power beam size is $21\arcsec$ ($11\arcsec$). We placed all CO spectra and line intensities on the main beam temperature scale ($T_{\rm mb}$), which is defined as $T_{\rm mb} = (F_{\rm eff}/B_{\rm eff})\times T_{\rm A}^*$. For the data reduction, we selected observations with good quality (obtained during satisfactory weather conditions and showing a flat baseline), averaged the spectra from individual scans of the source, and subtracted a constant continuum level.

\begin{figure}
\begin{center}
\includegraphics[width=0.52\textwidth]{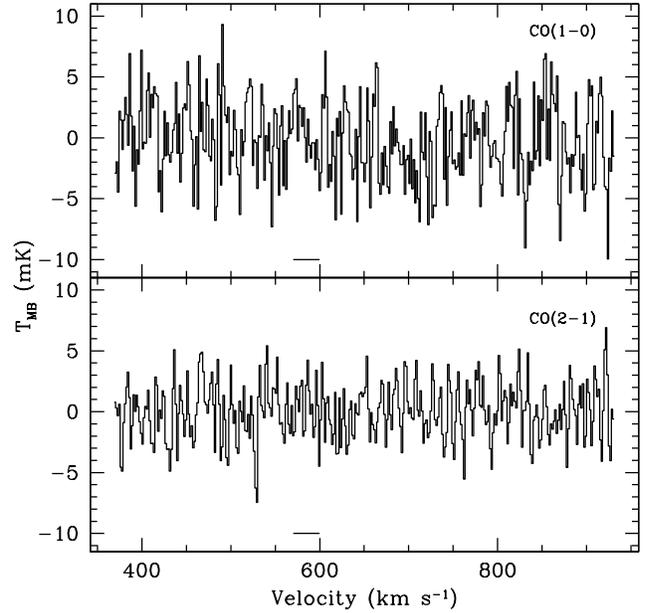}
\caption{CO(1--0) (top) and CO(2--1) (bottom) spectra for R2 in NGC~4625,
both with a channel width of $1.6 \, {\rm km \, s^{-1}}$. The horizontal line
indicates the expected location of the emission, based on the
H\,{\small I} velocity field from \citet{bush04} and assuming a line
width of $18.2 \, {\rm km \, s^{-1}}$.}
\label{spectrum}
\end{center}
\end{figure}

Figure~\ref{spectrum} shows the CO(1--0) and CO(2--1) spectra of R2 in NGC~4625. We did not detect emission in either transition. We calculated a 3$\sigma$ upper limit on the CO(1--0) and CO(2--1) line intensities with $I_{\rm CO} < 3 \, \sigma_{\rm rms} \, (\Delta V \, \delta v)^{1/2}$, where $\sigma_{\rm rms}$ is the noise in the spectrum, $\delta v$ is the spectrum channel width, and $\Delta V$ is the line width, which we assumed to be $18.2 \, {\rm km \, s^{-1}}$ to agree with \citet{braine07}. We measured an rms noise of $4.0 \, {\rm mK}$ with a channel width of $0.8 \, {\rm km \, s^{-1}}$ at CO(1--0) and $2.3 \, {\rm mK}$ with a channel width of $1.6 \, {\rm km \, s^{-1}}$ at CO(2--1).

For R4-R12 in NGC~4625, we calculated a shallow upper limit for the CO(2--1) line intensity by measuring the standard deviation of the average surface brightness in about ten emission-free regions in the HERACLES CO(2--1) integrated intensity map \citep{leroy09}. We do not quote an upper limit for R3 because it falls outside the HERACLES map. For the star forming regions in NGC~6946, M33, and NGC~4414, we used the deep CO(1--0) line intensities measured with the IRAM $30 \, {\rm m}$ telescope by \citet{braine07}, \citet{braine10}, and \citet{braine04}, respectively. For the non-detections, \citet{braine07} assumed a line width of $18.2 \, {\rm km \, s^{-1}}$ in calculating the 3$\sigma$ upper limits.

We used the following equation to calculate the molecular hydrogen surface density for the XUV disk star forming regions:
\begin{equation}
\label{eqn:sigma_H2}
\Sigma_{\rm H_{2}} [{\rm M}_{\odot} \, {\rm pc^{-2}}]= 3.2 \, I_{\rm CO(1-0)} \, {\rm cos}(i),
\end{equation}
where $I_{\rm CO(1-0)}$ is the CO(1--0) line intensity in ${\rm K \, km \, s^{-1}}$ (for R4-R12 in NGC~4625, we assumed $I_{\rm CO(2-1)}/I_{\rm CO(1-0)} = 0.7$), $i$ is the galaxy inclination \citep[$47\degr$ for NGC~4625, $33\degr$ for NGC~6946, $54\degr$ for M33, and $56\degr$ for NGC~4414;][]{schruba11,devaucouleurs91} and we do not include a correction for He. We estimated that the uncertainty on ${\rm log}(\Sigma_{\rm H_{2}})$ is $0.08 \, {\rm dex}$, based on the $\sim 15\%$ uncertainty due to the rms noise in the spectra and the $\sim 10\%$ uncertainty associated with flux calibration at CO(1--0).

Equation~\ref{eqn:sigma_H2} assumes the Milky Way CO-to-${\rm H_{2}}$ conversion factor of $X_{\rm CO} = 2 \times 10^{20} \, {\rm cm^{-2} \, (K \, km \, s^{-1})^{-1}}$. Both observations and theory indicate that $X_{\rm CO}$ is metallicity-dependent and increases sharply below $12 + {\rm log(O/H) = 8.2 - 8.4}$, although there are significant differences in the theoretical predictions for the slope of the $X_{\rm CO}$-metallicity relation at low-metallicity \citep{wolfire10,glover11,leroy11,narayanan12,bolatto13}. We selected the Milky Way CO-to-${\rm H_{2}}$ conversion factor because only two regions with deep CO data have metallicity measurements that we could use to calculate a metallicity-dependent $X_{\rm CO}$ (R2 in NGC~4625 has $12+{\rm log(O/H) = 8.5}$ and P3 in NGC~6946 has $12+{\rm log(O/H) = 8.1}$; see Table~\ref{tab:obs_prop2}). Furthermore, our choice is supported by the measurements of $X_{\rm CO}$ in NGC~4625 and NGC~6946 by \citet{sandstrom13}, who exploited the fact that dust traces the total gas column to solve for $X_{\rm CO}$ and the DGR using maps of the H\,{\small I} surface density, the CO integrated intensity, and the dust mass derived from spectral energy distribution fits to {\it Spitzer} and {\it Herschel} data. The radial profiles of $X_{\rm CO}$ in NGC~4625 and NGC~6946 extend to $2.5 \, r_{25}$ and $1.2 \, r_{25}$, respectively. \citet{sandstrom13} concluded that the average $X_{\rm CO}$ in NGC~6946 is $9 \times 10^{19} \, {\rm cm^{-2} \, (K \, km \, s^{-1})^{-1}}$. The value is low compared to the Milky Way value mainly because there is a deep depression of $X_{\rm CO}$ in the center of the galaxy. The profile flattens for radii larger than $0.5 \, r_{25}$, where $X_{\rm CO}$ for high-confidence regions (with $I_{\rm CO} > 1 \, {\rm K \, km \, s^{-1}}$) is primarily within a factor of two scatter of the Milky Way value. In NGC~4625, \citet{sandstrom13} do not quote an average $X_{\rm CO}$ because they have no high-confidence regions. The $X_{\rm CO}$ radial profile has few points and large uncertainties, but there is no clear evidence to support choosing a non-Milky Way value.

\subsection{Star Formation Rate Data and Measurements}
\label{sec:sfrs}
For NGC~4625 and NGC~6946, we estimated the SFR surface density with hybrid ${\rm H}\alpha + 24 \, \mu {\rm m}$ and ${\rm FUV} + 24 \, \mu {\rm m}$ calibrations. For M33 and NGC~4414, we only calculated the SFR surface density with the ${\rm FUV} + 24 \, \mu {\rm m}$ calibration because H$\alpha$ data was not available from surveys like the Local Volume Legacy Survey \citep[LVL;][]{kennicutt08} or the Spitzer Infrared Nearby Galaxies Survey \citep[SINGS;][]{kennicutt03}. We used the calibrations of \citet{leroy12}:
\begin{multline}
\label{eqn:SFR_Halpha}
\Sigma_{\rm SFR, H\alpha} [{\rm M}_{\odot} \, {\rm yr^{-1} \, kpc^{-2}}]=
(634 \, I_{\rm H\alpha} [{\rm erg \, s^{-1} \, cm^{-2} \, sr^{-1}}] \\
+ 0.9 \, \frac{I_{24}}{400 \, {\rm MJy \, sr^{-1}}}) \, {\rm cos}(i),
\end{multline}
\begin{multline}
\label{eqn:SFR_FUV}
\Sigma_{\rm SFR, FUV} [{\rm M}_{\odot} \, {\rm yr^{-1} \, kpc^{-2}}]= (8.1
\times 10^{-2} \, I_{\rm FUV} [{\rm MJy \, sr^{-1}}] \\ 
+ 1.3 \, \frac{I_{24}}{400 \, {\rm MJy \, sr^{-1}}}) \, {\rm cos}(i),
\end{multline}
where $I_{\rm H\alpha}$, $I_{\rm FUV}$, and $I_{24}$ are the H$\alpha$, FUV, and $24 \, \mu {\rm m}$ surface brightnesses within the $21\arcsec$-diameter aperture and $i$ is the galaxy inclination. $I_{\rm H\alpha}$ and $I_{\rm FUV}$ are corrected for Galactic extinction. The SFR calibrations are appropriate for regions containing continuous star formation, which requires $\sim 1 \, {\rm kpc}$-scale regions with $\Sigma_{\rm SFR} > 10^{-3} \, {\rm M_{\odot} \, yr^{-1} \, kpc^{-2}}$ in actively star forming galaxies \citep{leroy12}. At the distance of NGC~4625, NGC~6946, M33, and NGC~4414, $21\arcsec$ corresponds to $1 \, {\rm kpc}$, $0.6 \, {\rm kpc}$, $100 \, {\rm pc}$, and $1.8 \, {\rm kpc}$, respectively. While most of the apertures have large physical sizes, star formation is sparse in outer disks and the majority of our apertures contain only one star forming region with $\Sigma_{\rm SFR} < 10^{-3} \, {\rm M_{\odot} \, yr^{-1} \, kpc^{-2}}$. Because we used the \citet{leroy12} calibrations in these conditions, evolution of single star forming regions can affect the location on the Kennicutt-Schmidt law \citep{schruba10}. We discuss this further in Section~\ref{sec:discussion}.

We used the SFR calibrations that are appropriate when the $24 \, \mu {\rm m}$ data are not corrected for cirrus emission, which is the contribution to dust heating from non-star forming ($\ga 100 \, {\rm Myr}$) stellar populations. We included an uncertainty of $0.2 \, {\rm dex}$ on the SFR surface densities to account for the uncertainty due to the $24 \, \mu{\rm m}$ cirrus.

The SFRs we calculated based on H$\alpha + 24 \, {\rm \mu m}$ data and ${\rm FUV + 24 \, \mu m}$ data are consistent with each other in NGC~4625 and NGC~6946. The Kolmogorov-Smirnov test probability that the SFRs are drawn from the same population is 0.3.

In the following sections, we describe the H$\alpha$, FUV, and $24 \, \mu{\rm m}$ flux measurements that we used to calculate the SFR surface density. We also describe measurements of the H$\alpha$ and FUV fluxes within a $6\arcsec$-diameter aperture for NGC~4625 and NGC~6946. We used the $6\arcsec$-diameter fluxes, corrected for both Galactic and internal extinction, to estimate the ages of the star forming regions. Our age measurement assumes a single stellar population, so we used the smallest aperture possible given the GALEX FWHM (see also Section~\ref{sec:age}).

\subsubsection{H$\alpha$}
\label{sec:halpha}
For R2-R12 in NGC~4625, we used the IRAF\footnote{IRAF is distributed by the National Optical Astronomy Observatory, which is operated by the Association of Universities for Research in Astronomy (AURA) under cooperative agreement with the National Science Foundation.} task {\tt phot} to measure the H$\alpha$ flux from the LVL \citep{kennicutt08} image, which was obtained at the Steward Observatory $2.3 \, {\rm m}$ Bok telescope under seeing conditions of $\sim 1.6\arcsec$. \citet{kennicutt08} removed the stellar continuum (including stellar H$\alpha$ absorption) from the image by subtracting a scaled R-band image. Because R2 is on the edge of the image, we measured the flux within a $10\arcsec$-diameter aperture and used this value as a lower limit for the flux within the $21\arcsec$-diameter aperture.

We calculated the H$\alpha$ fluxes within $6\arcsec$- and $21\arcsec$-diameter apertures centered on the \citet{gildepaz07} regions. We did not subtract a sky background because the median sky value is consistent with zero. We used the formula in the LVL DR5 User's Guide\footnote{http://irsa.ipac.caltech.edu/data/SPITZER/LVL/ LVL\_DR5\_v5.pdf} to convert the flux in counts to ${\rm erg \, s^{-1} \, cm^{-2}}$. We applied an aperture correction of 1.42 to the fluxes within the $6\arcsec$-diameter aperture, but applied no aperture correction to the fluxes within the $21\arcsec$-diameter aperture because the total flux of a point source is contained within that aperture. We corrected for [N\,{\small II}] $\lambda \lambda 6548, 6584$ emission in the image using the [N\,{\small II}]$\lambda 6584/{\rm H}\alpha$ line ratio measured for each region by \citet{gildepaz07}, who used a spectrum obtained with a $1.5\arcsec$ slit on the Palomar Observatory $5 \, {\rm m}$ telescope (for R3, we used the average ratio from the remaining regions). We assumed the theoretical value of [N\,{\small II}]$\lambda 6548/6584$ of 0.33 to correct for both [N\,{\small II}] lines \citep{osterbrock89}. For the flux within the $21\arcsec$-diameter aperture, we corrected for Galactic extinction using $A_{V}$ values from \citet{schlafly11} and the \citet{odonnell94} extinction law to calculate the extinction at the effective wavelength of the H$\alpha$ filter ($6585 \, {\rm \AA}$), assuming $R_{V} = 3.1$. For the flux within the 6$\arcsec$-diameter aperture, we corrected for Galactic plus internal extinction using the E(B-V) values derived in \citet{gildepaz07} from the Balmer decrement and the \citet{odonnell94} extinction law.

We assigned an uncertainty of 20\% to the H$\alpha$ fluxes within the $6\arcsec$- and $21\arcsec$-diameter apertures. We included an uncertainty of 14\% based on the uncertainty quoted for the integrated H$\alpha$+[N\,{\small II}] flux of NGC~4625 in \citet{kennicutt08}. The primary contribution to this uncertainty is from the continuum removal, but \citet{kennicutt08} also included the uncertainty in the flat-fielding and the photometric zero point. We also measured and included a $\sim 13\%$ uncertainty due to the large rms of the sky background, and a $\sim 10\%$ uncertainty due to the fact that the aperture containing the total point-source flux is uncertain.

For the star forming regions in NGC~6946, we measured the fluxes from an H$\alpha$ image obtained by \citet{ferguson98b} with the Kitt Peak National Observatory $0.9 \, {\rm m}$ with $\sim 2.0\arcsec$ seeing. \citet{ferguson98b} also removed the stellar continuum by subtracting a scaled R-band image. We followed the same flux measurement procedures as for NGC~4625, except for the following differences. First, P4 is not detected and we calculated the 3$\sigma$ upper limit based on the image noise and the area of the aperture. Second, our aperture centering method depended on how we used the fluxes. We used the fluxes within the $21\arcsec$-diameter aperture for the analysis on the Kennicutt-Schmidt law. The SFR and CO apertures must be matched for this analysis, so we centered the H$\alpha$ apertures on the coordinates of the CO measurements from \citet{braine07}, which are all near H\,{\small II} regions, but not necessarily centered on the H$\alpha$ or FUV peak. We used the fluxes within the $6\arcsec$-diameter aperture to estimate the age of the individual star forming regions. Therefore, we centered on the H$\alpha$ peak closest to the \citet{braine07} coordinates. This was a clear choice in every case but P7, which has two H$\alpha$ peaks within $7\arcsec$ of the \citet{braine07} coordinates; we chose the nearest peak.

Third, we applied an aperture correction of 1.19 to the fluxes of the detected regions within the $6\arcsec$-diameter aperture (again no correction is necessary for the $21\arcsec$-diameter aperture). Fourth, we applied a single factor of 0.903 to correct for [N\,{\small II}] $\lambda \lambda 6548, 6584$ emission in all the NGC~6946 regions. This factor is based on the measured line intensities for P3 \citep[Region C in][]{ferguson98a}. Based on the range of correction factors measured for the eleven NGC~4625 regions in \citet{gildepaz07} and the expected values from the model of \citet{dopita06}, this single correction factor is likely correct to within $5 - 10 \%$. Finally, we corrected the $6\arcsec$-diameter fluxes for internal extinction using the $A_{\rm FUV}$ values derived from the \citet{munoz09} radial profile (see Section~\ref{sec:galex}) and assuming $A_{\rm FUV} = 1.8 \, A_{\rm H\alpha}$, where both extinctions refer to the values due to only internal extinction \citep{calzetti01}. This extinction relation includes higher dust obscuration towards the youngest massive stars because older stars have left or destroyed their natal clouds (note that the Cardelli et al. 1989 extinction law for the Milky Way with $R_{V} = 3.1$ gives $A_{\rm FUV} = 3.2 \, A_{\rm H\alpha}$). The \citet{calzetti01} extinction relation was originally calculated for starbursts, but also applies in dwarfs \citep{lee09}, and a similar relation applies in M83 \citep{boissier05}. Note that this choice of extinction correction affects our results. We discuss this further in Section~\ref{sec:age}.

The uncertainty on the NGC~6946 fluxes that are corrected for Galactic extinction is the same as for the NGC~4625 regions ($20\%$), even including the additional uncertainty in the [N\,{\small II}] correction. The uncertainty on the fluxes that are corrected for Galactic plus internal extinction is $30\%$, which accounts for the average uncertainty in the $A_{\rm FUV}$ values estimated from the \citet{munoz09} radial profile.

The fluxes we calculated within the $21\arcsec$-diameter aperture are about a factor of two larger than the values presented in \citet{braine07}. The most likely reasons for the discrepancy are the sky subtraction and the Galactic extinction correction. We chose to include the diffuse H$\alpha$ emission in our apertures, which is common in extragalactic star formation studies (e.g., Leroy et al.\ 2012; but see also, e.g., Blanc et al.\ 2009). Our fluxes would be 40\% smaller if we subtracted the median sky background in an annulus surrounding each aperture. There would be a 20\% difference in our fluxes if we had used the Galactic extinction values from \citet{schlegel98}, albeit in the wrong direction to explain the discrepancy. However, this illustrates the magnitude of discrepancy possible in correcting for Galactic extinction.

\subsubsection{{\it GALEX} FUV}
\label{sec:galex}
We measured the FUV fluxes for the star forming regions in NGC~4625, M33, and NGC~4414 from the {\it GALEX} Nearby Galaxy Survey (NGS) images from \citet{gildepaz07b}. For NGC~6946, we used the NGS image that was processed by the {\it GALEX} pipeline. All images are available on the NASA Mikulski Archive for Space Telescopes. The effective wavelength is $1516 \, {\rm \AA}$ and the FWHM is $\sim 4.5\arcsec$. For the detected regions, we used the IRAF task {\tt phot} to measure the FUV flux within $21\arcsec$-diameter apertures. For the NGC~4625 and NGC~6946 regions, we also measured the FUV flux within $6\arcsec$-diameter apertures and we used the same aperture centers as we used for the H$\alpha$ measurements (see Section~\ref{sec:halpha}). For M33 and NGC~4414, we centered the apertures on the coordinates of the CO measurements from \citet{braine10} and \citet{braine04}, respectively. We subtracted the median sky background from each pixel value. We applied an aperture correction, calculated from point sources in the NGC~6946 image, of 1.72 (1.04) to the fluxes within the $6\arcsec$ ($21\arcsec$) diameter aperture. We corrected for Galactic extinction using the $A_{V}$ values from \citet{schlafly11} and the extinction law of \citet{cardelli89} with $R_{V}= 3.1$. For the undetected regions in NGC~6946, we quote 3$\sigma$ upper limits, which we calculated using the image noise and the aperture area.

For the $6\arcsec$-diameter fluxes, we also corrected for the internal FUV extinction ($A_{\rm FUV}$). In NGC~4625, we assumed $A_{\rm FUV} = 1.8 \, A_{\rm H\alpha}$, where both extinctions are due to only internal extinction \citep[][see also Section~\ref{sec:halpha}]{calzetti01}. We calculated $A_{\rm H\alpha}$ based on the ratio of the H$\alpha$ flux corrected for Galactic and internal extinction \citep[using the ${\rm E(B-V)}$ values from][]{gildepaz07} to the H$\alpha$ flux corrected for only Galactic extinction \citep[using $A_{V}$ from][]{schlafly11}. For NGC~6946, we used the \citet{munoz09} $A_{\rm FUV}$ radial profile for the galaxy, which the authors computed based on the relationship between the internal FUV extinction and the total-IR (TIR)-to-FUV luminosity ratio. We used the results based on the star formation history-independent relationship of Buat et al. (2005) and used a linear interpolation between $A_{\rm FUV}$ estimates in the radial profile. Note that the extinction based on a radial average may be an underestimate for star forming regions, which likely have extinctions at the high end of the distribution.

We calculated the fractional uncertainty on the FUV fluxes to be 11\% for all fluxes for NGC~4625 and for the NGC~6946, M33, and NGC~4414 fluxes that are corrected for only Galactic extinction. This includes the average uncertainty due to photometric repeatability and photon noise \citep[10\%;][]{morrissey07} and an uncertainty in the background level (5\%). For the NGC~6946 fluxes that are additionally corrected for internal extinction, we calculated the uncertainty to be 36\%. This includes the uncertainties above plus the average uncertainty due to the internal extinction correction based on the \citet{munoz09} radial $A_{\rm FUV}$ profile (the $A_{\rm FUV}$ uncertainty is larger at larger radius and leads to flux uncertainty from 8\% - 46\%).

\subsubsection{{\it Spitzer} MIPS $24 \, \mu {\rm m}$}
For NGC~4625 and NGC~6946, we measured the $24 \, \mu {\rm m}$ surface brightness of each star forming region on {\it Spitzer Space Telescope} Multiband Imaging Photometer for Spitzer (MIPS) images from the fifth and final data release of SINGS \citep{kennicutt03}. We used the M33 image from LVL \citep{dale09} and the NGC~4414 post-Basic Calibrated Data (PBCD) image available on the NASA/IPAC Infrared Science Archive (PI: G. Fazio).

One region in NGC~4625 (R6) and three regions in NGC~6946 (P7, P2, and P8) are detected. There is a source near NGC~4625 R2, but it is most likely a background galaxy. It is unlikely that the source is a star forming region in NGC~4625 because of the lack of associated emission in the FUV, optical, or H$\alpha$. It is unlikely that it is a foreground star because it does not show point-like NUV emission \citep{bigiel08}. The M33 image has significant contamination from foreground and background sources. We compared the $24 \, \mu {\rm m}$, FUV, NUV, and optical images and concluded that the star forming regions are not detected at $24 \, \mu {\rm m}$, although four of the six regions are contaminated by background galaxies or single variable stars in M33 \citep{hartman06}. For the non-detections, we used $3\sigma$ upper limits, which we measured as three times the standard deviation of the average surface brightness in many emission-free regions.

For the detections, we measured the average surface brightness in the $21\arcsec$-diameter aperture on the original resolution images (${\rm FWHM} \sim 6\arcsec$) with the IRAF task {\tt phot}. We did not subtract a sky background level for NGC~4625 because the median background value is consistent with zero. The NGC~6946 image has a large-scale, variable background (described in the SINGS fifth data release document\footnote{http://irsa.ipac.caltech.edu/data/SPITZER/SINGS/doc/ sings\_fifth\_delivery\_v2.pdf}). Therefore, we subtracted the local background level, which we measured in an annulus between $31.5\arcsec$ and $45\arcsec$ from the center of each region. We measured an aperture correction of 1.19 for the $21\arcsec$-diameter aperture and applied this correction to the surface brightness measurements. We estimated an uncertainty of 10\% on the surface brightness values in NGC~4625 and 20\% on the surface brightness values in NGC~6946. These uncertainties include the suggested values from the MIPS Instrument Handbook\footnote{http://irsa.ipac.caltech.edu/data/SPITZER/docs/mips/ mipsinstrumenthandbook/} of 4\% calibration uncertainty and 5\% uncertainty in the aperture correction, as well as an uncertainty due to the sky background, including the uncertainty introduced because of the large-scale, variable background in NGC~6946. 

\begin{figure}
\begin{center}
\includegraphics[width=0.5\textwidth]{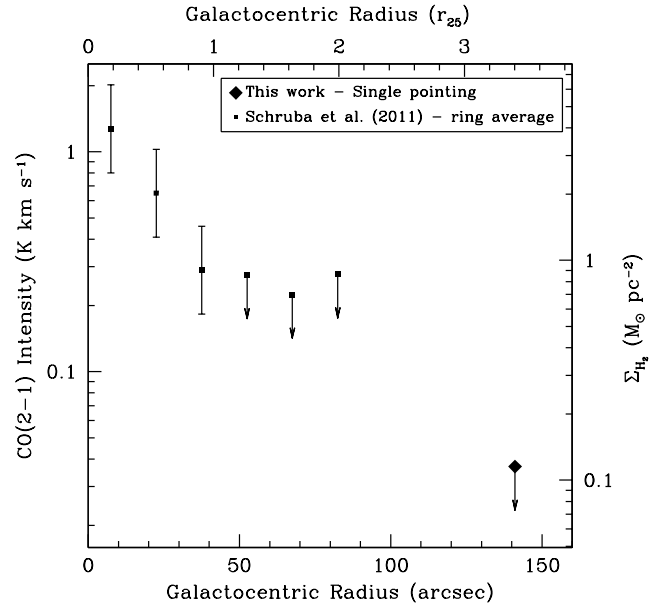}
\caption{CO(2--1) radial profile in NGC~4625, where we used $r_{25} =
  41.4\arcsec$ \citep{schruba11} for the top axis and $I_{\rm
    CO(2-1)}/I_{\rm CO(1-0)} = 0.7$ to convert the CO(2--1) intensity
  to $\Sigma_{\rm H_{2}}$ for the right axis. This demonstrates the
  deep sensitivity of our $3\sigma$ CO upper limit (diamond) relative to
  radial averages within the optical disk (squares) from
  \citet{schruba11}.}
\label{radial_profile}
\end{center}
\end{figure}

\begin{figure*}
\begin{center}
\includegraphics[width=\textwidth]{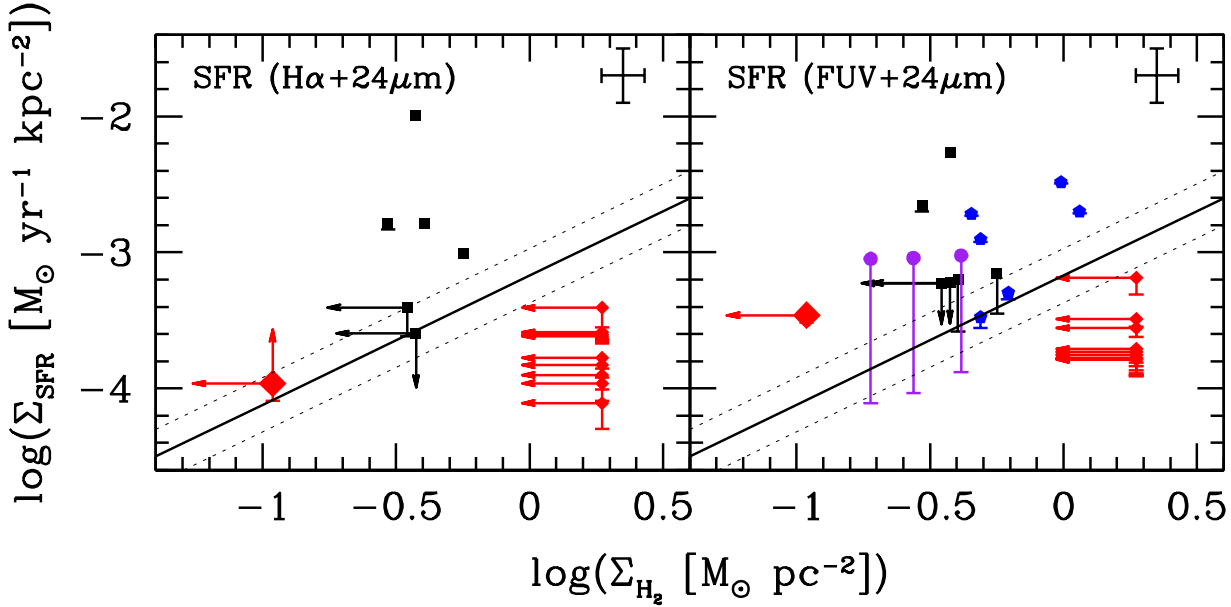}
\caption{SFR surface density versus molecular hydrogen surface density for XUV disk star forming regions within the $21\arcsec$-diameter aperture ($1 \, {\rm kpc}$ for NCC~4625, $0.6 \, {\rm kpc}$ for NGC~6946, $100 \, {\rm pc}$ for M33, and $1.8 \, {\rm kpc}$ for NGC~4414). The SFR in the left panel is based on H$\alpha$ and $24 \, \mu {\rm m}$ data while the SFR in the right panel is based on FUV and $24 \, \mu {\rm m}$ data following the prescriptions of \citet{leroy12}. The large red diamond represents R2 in NGC~4625, where we used the deep CO(1--0) upper limit measured in this work. The small red diamonds represent R4-R12 in NGC~4625, with CO(2--1) upper limits from the HERACLES integrated intensity map \citep{leroy09} and assuming $I_{\rm CO(2-1)}/I_{\rm CO(1-0)} = 0.7$ \citep{schruba11}. The black squares, blue pentagons, and purple circles represent star forming regions in NGC~6946, M33, and NGC~4414, respectively, with CO(1--0) measurements or upper limits from \citet{braine07}, \citet{braine10}, and \citet{braine04}. The range of possible $\Sigma_{\rm SFR}$ values is shown for regions with a detection in one SFR tracer and a non-detection in the other SFR tracer. A typical error bar is in the upper right corner. The solid line represents the molecular-hydrogen Kennicutt-Schmidt law within the optical disk of 30 spiral galaxies and dotted lines show the $1\sigma$ scatter \citep{leroy13}. The XUV disk star forming regions are consistent with the same molecular-hydrogen Kennicutt-Schmidt law that applies within the optical disk.}
\label{schmidt}
\end{center}
\end{figure*}

\section{Results}
\label{sec:results}
\subsection{\boldmath{Deep CO Upper Limit at $3.4 \, {r_{25}}$ in NGC~4625}}
Figure~\ref{radial_profile} shows our 3$\sigma$ upper limit on the CO(2--1) line intensity for R2 at $r=3.4 \, r_{25}$ in NGC~4625 (diamond) relative to the radial profile from \citet{schruba11}. \citet{schruba11} achieved a sensitivity of $I_{\rm CO(2-1)} \sim 0.3 \, {\rm K \, km \, s^{-1}}$ ($\Sigma_{\rm H_{2}} \sim 1 \, {\rm M}_{\odot} \, {\rm pc^{-2}}$) by stacking the spectra (with $13\arcsec$-FWHM resolution) within $15\arcsec$-wide tilted rings. Our deep CO exposure on R2 provided a 3$\sigma$ upper limit of $I_{\rm CO(2-1)} < 0.04 \, {\rm K \, km \, s^{-1}}$ ($I_{\rm CO(1-0)} < 0.05 \, {\rm K \, km \, s^{-1}}$; $\Sigma_{\rm H_{2}} < 0.11 \, {\rm M}_{\odot} \, {\rm pc^{-2}}$), a factor of seven deeper sensitivity than was reached in the \citet{schruba11} profile. Our upper limit on the CO line intensity is also deeper than the published measurements and limits for other outer disk star forming regions: \citet{braine04}, \citet{braine07}, \citet{braine10}, and \citet{dessauges14}, all of which used the IRAM $30 \, {\rm m}$ telescope. The deepest previously published values are $I_{\rm CO(1-0)} = 0.11 \, {\rm K \, km \, s^{-1}}$ \citep{braine04,braine07,dessauges14} and $I_{\rm CO(2-1)} < 0.06 \, {\rm K \, km \, s^{-1}}$ \citep{dessauges14}. R2 in NGC~4625 is also at a larger radius relative to $r_{25}$ than the published regions, all of which are at $r < 2 \, r_{25}$.

\subsection{Molecular-Hydrogen Kennicutt-Schmidt Law for Star Forming Regions in XUV Disks}
\label{sec:ks}
In Figure~\ref{schmidt}, we compared the location of XUV disk star forming regions with deep CO measurements or upper limits (large red diamond and small black squares, blue pentagons, and purple circles) to the \citet{leroy13} fit to the molecular-hydrogen Kennicutt-Schmidt law within the optical disk of 30 spiral galaxies (solid line, with $1\sigma$ scatter shown as dotted lines; we used the fit that is appropriate when the $24 \, \mu{\rm m}$ data are not corrected for cirrus emission and we adjusted the relation so it is appropriate for $\Sigma_{\rm H_{2}}$ values that are not corrected for helium). We also included NGC~4625 R4-R12 from \citet{gildepaz07}, which have shallow CO(2--1) upper limits from HERACLES \citep[small red diamonds;][]{leroy09}.

When we computed the SFRs based on H$\alpha + 24 \mu {\rm m}$ data or ${\rm FUV} + 24 \mu {\rm m}$ data (the left and right panels of Figure~\ref{schmidt}, respectively), $76-81\%$ of all the regions, $57-63\%$ of the regions with deep CO observations, and $25-62\%$ of the regions with CO detections are or could be within the $1\sigma$ scatter of the \citet{leroy13} fit. The detections tend to be offset to larger $\Sigma_{\rm SFR}$ at a given $\Sigma_{\rm H_{2}}$ compared to the fit for the optical disk. This offset to high SFE is consistent with single star forming regions that are relatively evolved, in that they may represent clusters with relatively high SFRs that have exhausted or expelled a fraction of their molecular gas, as proposed  by \citet{schruba10} for some regions in M33. Therefore, while most of the regions are consistent with the molecular-hydrogen Kennicutt-Schmidt law, the largest outliers may be due to evolutionary effects.

\subsection{Age Estimates for Star Forming Regions in XUV Disks}
\label{sec:age}
\begin{figure}
\begin{center}
\includegraphics[width=0.5\textwidth]{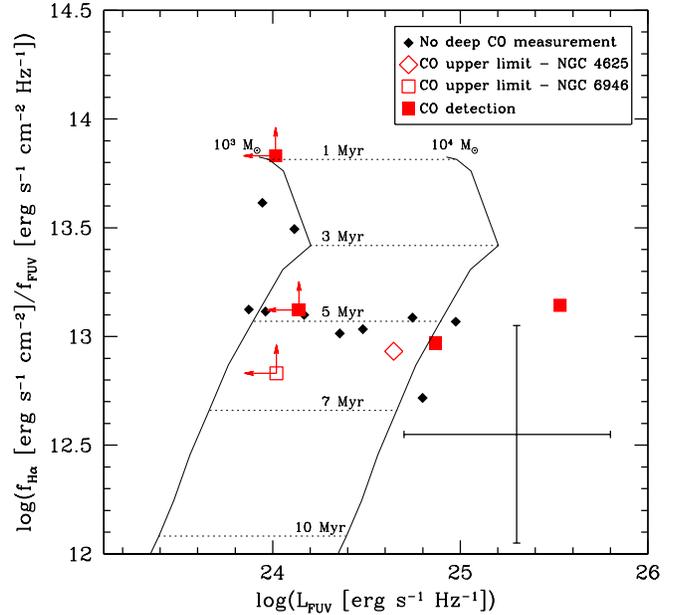}
\caption{Ratio of H$\alpha$ to FUV flux versus FUV luminosity for XUV disk star forming regions in NGC~4625 and NGC~6946. Open red symbols represent regions with a deep CO upper limit (the diamond refers to R2 in NGC~4625 and the square refers to P5 in NGC~6946). The filled red squares represent regions with a CO detection in NGC~6946 from Braine et al. (2007). Black diamonds represent R3-R12 in NGC~4625, which have no deep CO measurement attempted. The lines represent Starburst99 models for the evolution of the H$\alpha$ and FUV emission of star clusters with masses of $10^{3}$ and $10^{4} \, {\rm M}_{\odot}$. The error bars include two potential sources of systematic uncertainty: aperture size (a smaller aperture could give H$\alpha$-to-FUV flux ratios larger by a factor of three and FUV luminosities smaller by a factor of four) and the internal extinction correction (use of the \citet{cardelli89} extinction law would result in H$\alpha$-to-FUV flux ratios smaller by a factor of three and FUV luminosities larger by a factor of three). We do not find any trend between CO detection rate and age. Deeper CO data for more regions are needed to test our prediction that all the regions plotted should have molecular gas because they are young compared to the typical lifetime of a molecular cloud \citep[$\sim 20-30 \, {\rm Myr}$;][]{mckee07}.}
\label{age}
\end{center}
\end{figure}

We estimated that the ages of the star forming regions in the XUV disks of NGC~4625 and NGC~6946 are $1 - 7 \, {\rm Myr}$, based on the ratio of the H$\alpha$ flux to the FUV flux, both of which are corrected for Galactic and internal extinction \citep[Figure~\ref{age}; see also][]{goddard11}. This age estimate requires that we measured the flux ratio of a single stellar population that underwent a burst of star formation. We therefore used the smallest aperture appropriate for the {\it GALEX} FWHM: the $6\arcsec$-diameter aperture. We compared our H$\alpha$-to-FUV flux ratio measurements to Starburst99 models \citep{leitherer99} for the evolution of $10^{3} \, {\rm M}_{\odot}$ and $10^{4} \, {\rm M}_{\odot}$ clusters (solid lines in Figure~\ref{age}) with the metallicity of the Large Magellanic Cloud ($12+{\rm log(O/H)=8.4}$; $Z=0.008$). We chose this metallicity because it is the closest option offered in the code relative to the average metallicity of the star forming regions we studied ($12+{\rm log(O/H) = 8.3}$; see Table~\ref{tab:obs_prop2}). We used the default Starburst99 IMF, which has an upper boundary of $100 \, M_{\odot}$. The age range we measured is consistent with the ages first reported by \citet{goddard11}, with overlap in the regions we studied in NGC~4625. \citet{goddard11} cautioned that the ages in this plot may be spuriously old if stochasticity in the star formation process is relevant in the regions. 

To put the age estimates in the context of the presence or absence of molecular gas, black diamonds in Figure~\ref{age} represent the star forming regions where no deep CO measurement has been attempted, open red symbols represent regions with a deep CO upper limit, and filled red symbols represent regions with a CO detection from \citet{braine07}. We found no trend between age and the CO detection rate. In Section~\ref{sec:discussion}, we discuss this result relative to our expectation that all the regions we studied should contain molecular gas because they are young compared to the typical destruction time of a molecular cloud \citep[$20-30 \, {\rm Myr}$;][]{mckee07,kawamura09}.

We estimated that the uncertainty on the H$\alpha$-to-FUV flux ratio is $\pm 0.5 \, {\rm dex}$ and the uncertainty on the FUV luminosity is $\substack{+0.5 \\ -0.6} \, {\rm dex}$, based on the potential for systematic uncertainty due to the aperture size and the internal extinction correction (described in more detail below). The flux ratio uncertainty leads to an uncertainty of about $2 \, {\rm Myr}$ in the age estimate. This does not include the systematic uncertainty due to stochasticity in the star formation process (mentioned above) or the assumed IMF.

The first potential source of systematic uncertainty is due to the aperture size, which affects the H$\alpha$-to-FUV flux ratio and the FUV luminosity in NGC~4625 because there is abundant diffuse FUV emission in the XUV disk while H$\alpha$ is mainly from concentrated regions. The H$\alpha$-to-FUV flux ratio within the 6$\arcsec$-diameter aperture is on average a factor of 2.9 larger than the value within the 21$\arcsec$-diameter aperture. The FUV luminosity within the 6$\arcsec$-diameter aperture is on average a factor of 4.2 smaller than the value within the 21$\arcsec$-diameter aperture. If we were not limited by the angular resolution of the FUV data, we would have used an aperture smaller than $6\arcsec$ ($280 \, {\rm pc}$ in NGC~4625 and $170 \ {\rm pc}$ in NGC~6946) for our flux ratio measurements because our age estimate assumes a single stellar population. The H$\alpha$ luminosities of the regions in NGC~4625 are consistent with gas ionized by a single O star \citep{gildepaz07,goddard11} and typical H\,{\small II} regions that are ionized by a single star have diameters less than $30 \, {\rm pc}$ \citep{whitmore11}. Based on the radial profile of the H$\alpha$-to-FUV flux ratio between a radius of $3\arcsec$ and 10.5$\arcsec$, we expect the flux ratio to be larger and ages younger within smaller apertures. We estimated that the H$\alpha$-to-FUV flux ratio in $\sim 30 \, {\rm pc}$-diameter regions could be approximately a factor of three larger than our nominal values, based on the comparison between the flux ratio within the 6$\arcsec$- and 21$\arcsec$-diameter aperture. We also estimated that the FUV luminosity in $\sim 30 \, {\rm pc}$-diameter regions could be approximately a factor of four smaller than our nominal values because of the prevalent extended emission. More precise estimates of the FUV luminosity, H$\alpha$-to-FUV flux ratio, and age would be possible with higher resolution FUV data. Note that aperture size does not uniformly affect all regions or all galaxies; the two regions with FUV detections in NGC~6946 show relatively flat H$\alpha$-to-FUV radial profiles because there is less diffuse FUV emission than in NGC~4625.

The second potential source of systematic uncertainty is due to the internal extinction correction.  We used the \citet{calzetti01} relation $A_{\rm FUV} = 1.8 \, A_{\rm H\alpha}$. If we instead applied the \citet{cardelli89} extinction law ($A_{\rm FUV} = 3.2 \, A_{\rm H\alpha}$), the NGC~4625 H$\alpha$-to-FUV flux ratios would be on average smaller by a factor of three and the FUV luminosities larger by a factor of three while the NGC~6946 flux ratios would be smaller by a factor of 1.2 on average. \citet{goddard10,goddard11} corrected for internal extinction using an azimuthally-averaged $A_{\rm FUV}$-(FUV-NUV) color relation \citep[note that starburst and normal galaxies adhere to different $A_{\rm FUV}$-(FUV-NUV) relationships and this type of relation is not reliable in dwarfs;][]{lee09}. They applied no extinction correction beyond the radius where the FUV-NUV color averages to zero and noted that outer disk star forming regions are often beyond this radius. If we did not apply any internal extinction correction, our H$\alpha$-to-FUV flux ratios would be on average larger by a factor of 1.7 and the FUV luminosities smaller by a factor of 3.8. We still found no trend between age and the CO detection rate if we corrected for internal extinction using the \citet{cardelli89} extinction law or if we applied no internal extinction correction.

\section{Discussion}
\label{sec:discussion}

\subsection{The Molecular-Hydrogen Kennicutt-Schmidt Law in XUV Disks: Evolutionary Effects and Comparison to Other Work}
In Section~\ref{sec:ks}, we concluded that the XUV disk star forming regions in Figure~\ref{schmidt} are in general consistent with the molecular-hydrogen Kennicutt-Schmidt law that applies within the optical disk, although there are significant outliers. This statement applies for SFRs calculated using either H$\alpha + 24 \, \mu{\rm m}$ or ${\rm FUV} + 24 \, \mu {\rm m}$ data. We also concluded that the distributions of SFRs based on H$\alpha + 24 \, \mu{\rm m}$ and ${\rm FUV} + 24 \, \mu {\rm m}$ are consistent with each other in NGC~4625 and NGC~6946 (the Kolmogorov-Smirnov test probability that the distributions are drawn from the same population is 0.3; Section~\ref{sec:sfrs}). One may have expected the H$\alpha$-based SFRs to be lower than the FUV-based SFRs because the H$\alpha$-to-FUV flux ratio in XUV disk star forming regions is lower by up to a factor of 1.6 compared to regions in the optical disk \citep{goddard10}. This discrepancy between our expectations and the observations may be due to our selection of H$\alpha$-bright regions in NGC~4625 and NGC~6946 combined with the fact that the SFRs both include $24 \, \mu {\rm m}$ emission, which is less sensitive to age than H$\alpha$.

We also concluded in Section~\ref{sec:ks} that the XUV disk star forming regions with CO detections tend to have high molecular gas SFEs (i.e., large $\Sigma_{\rm SFR}$ at a given $\Sigma_{\rm H_{2}}$) relative to values in typical inner disks. The high molecular gas SFEs could be due to evolutionary effects in our H$\alpha$- or FUV-selected regions. \citet{schruba10} studied the inner disk of M33 and concluded that scatter in the Kennicutt-Schmidt law increases when the aperture contains few star forming regions (${\rm diameter} < 300 \, {\rm pc}$) because the regions could be in any evolutionary state between a molecular cloud with little star formation and a relatively mature cluster with little remaining molecular gas. The scatter in the Kennicutt-Schmidt law decreases with larger aperture size because one averages over the evolutionary stages of multiple regions. Evolutionary effects may be important in XUV disk star forming regions because star formation is so sparse in outer disks that our aperture often only contains one star forming region even though most of the apertures have large physical sizes ($1 \, {\rm kpc}$ in NGC~4625, $0.6 \, {\rm kpc}$ in NGC~6946, $100 \, {\rm pc}$ in M33, and $1.8 \, {\rm kpc}$ in NGC~4414).

\citet{kruijssen14} developed an analytic expression for the aperture size below which the Kennicutt-Schmidt law breaks down due to large scatter and evaluated the expression for an idealized disk galaxy, assuming H$\alpha$ as the SFR tracer. They concluded that evolutionary effects set the minimum aperture size for the majority of the optical disk, in agreement with \citet{schruba10}. In the outer disk of their model galaxy where the SFR is low, the aperture size is set by the incomplete population of the IMF with massive stars. However, the underproduction of massive stars can only lead to underestimated SFRs. Therefore, evolutionary effects are more likely the cause of the high molecular gas SFEs in the XUV regions with CO detections.

In contrast to the high molecular gas SFEs that we derived for the regions with CO detections in NGC~6946, M33, and NGC~4414, \citet{dessauges14} concluded that their two CO detections in M63 at $r = 1.36 \, {r_{25}}$ have low molecular gas SFEs. Evolutionary effects may be less important for the regions studied by \citet{dessauges14} because their pointings cover a complex of FUV knots that may contain multiple star forming regions, although the lack of H$\alpha$ emission seems to indicate that the complex is uniformly older than $10 \, {\rm Myr}$. \citet{dessauges14} noted that the outer disk warp in M63 or high turbulence may lead to the lower SFEs that they observed.  To determine the range of SFEs possible in XUV disks, it will be useful to compile CO measurements for more regions with a variety of H$\alpha$-to-FUV flux ratios.

\subsection{The Presence of Molecular Gas as an Independent Age Probe}
In Section~\ref{sec:age}, we estimated ages for the XUV disk star forming regions in Figure~\ref{age} to be $1-7 \, {\rm Myr}$ based on the H$\alpha$-to-FUV flux ratios and we noted that these ages could be overestimated if stochastic sampling of the IMF is important. An additional -- albeit coarse -- probe of the age is the presence or absence of molecular gas, because star forming regions evolve from molecular clouds with little star formation to clusters that have exhausted or expelled the gas. We hypothesized that molecular gas should be present in all the regions in Figure~\ref{age} because molecular clouds start forming stars fast and live for $20-30 \, {\rm Myr}$ \citep[although the lifetime estimates are uncertain;][]{mckee07,kawamura09,dobbs13} while all our regions are younger than $7 \, {\rm Myr}$. However, we did not find any trend between the age based on the H$\alpha$-to-FUV flux ratio and the CO detection rate. Our current data are insufficient to fully test our hypothesis. We can make progress in comparing ages based on the H$\alpha$-to-FUV flux ratio versus the presence or absence of molecular gas in the low-density regime with deeper CO observations of more star forming regions and higher resolution FUV data.

A deep CO map of a large ($\ga {\rm kpc}$) region in an XUV disk could also help break the degeneracy between age and stochastic sampling of the IMF as the explanation for low H$\alpha$-to-FUV flux ratios in XUV disks. If star forming regions are only aging, we expect $2-3$ molecular clouds per H\,{\small II} region and $0.2-0.3$ molecular clouds per FUV-bright region, based on the relative lifetimes of molecular clouds that are capable of forming massive stars ($\sim 20-30 \, {\rm Myr}$), H\,{\small II} regions ($\sim 10 \, {\rm Myr}$), and the FUV-emitting phase ($\sim 100 \, {\rm Myr}$). If stochastic sampling of the IMF is important, some of the regions could appear old based on the H$\alpha$-to-FUV flux ratio but in reality be younger than $10 \, {\rm Myr}$. These regions would likely still contain molecular gas. Therefore, the map would contain more than $2-3$ molecular clouds per H\,{\small II} region. This test can be carried out in the near future, as the Atacama Large Millimeter/submillimeter Array has the sensitivity to map large regions of XUV disks at about the resolution of molecular clouds and detect CO emission from the least massive clouds that are capable of producing an O or B0 star \citep[$\sim 10^{4} \, {\rm M}_{\odot}$;][]{kennicutt12, koda12}.

It is not yet clear which conditions maximize the likelihood of detecting CO in XUV disks. We did not detect CO in a young region with H$\alpha$ emission in NGC~4625 while \citet{braine07} detected CO in regions with similar H$\alpha$-to-FUV flux ratios, SFR surface densities, and ages. High H\,{\small I} surface density could indicate regions that are partially molecular. The \citet{dessauges14} regions have H\,{\small I} surface densities ($2 \, {\rm M}_{\odot} \, {\rm pc^{-2}} \la \Sigma_{\rm HI} \la 4 \, {\rm M}_{\odot} \, {\rm pc^{-2}}$, not including He) that are similar to inner disk regions in M63 with CO detections. However, they only detected CO in two out of twelve XUV disk regions. In NGC~4625, the H\,{\small I} surface density at the largest radius ($r \sim 2 \, r_{25}$) in the \citet{schruba11} radial profile is $\Sigma_{\rm HI} \sim 2 \, {\rm M}_{\odot} \, {\rm pc}^{-2}$, which is less than the H\,{\small I} surface density in regions where CO is detected ($r<r_{25}$ with $\Sigma_{\rm HI} \sim 4 \, {\rm M}_{\odot} \, {\rm pc}^{-2}$). CO detectability in XUV disks may be correlated with the radius relative to $r_{25}$, stellar surface density, mid-plane pressure, and orbital time because these parameters are generally correlated with molecular fraction ($\Sigma_{\rm H_{2}}/\Sigma_{\rm HI}$; Wong \& Blitz 2002; Blitz \& Rosolowsky 2006; Leroy et al.\ 2008). Finally, the CO detection rate in XUV disks may be lower for low-mass galaxies because they tend to have lower metallicity \citep{tremonti04} and the CO luminosity decreases for a given ${\rm H_{2}}$ mass at low metallicity \citep[e.g.,][]{wolfire10,leroy11,bolatto13}.

\section{Conclusions}
We measured an upper limit on the CO intensity in one star forming region at $r=3.4 \, r_{25}$ in the XUV disk of NGC~4625. We measured the SFR surface density (with H$\alpha$, FUV, and $24 \, \mu {\rm m}$ data) for this region plus fifteen star forming regions in the XUV or outer disks of NGC~6946, M33, and NGC~4414 that have deep CO measurements or upper limits from the literature, and ten additional star forming regions in NGC~4625 with shallow CO upper limits from the HERACLES survey \citep{leroy09}. Our two main results based on these measurements are the following: \\

\noindent 1. The XUV disk star forming regions are in general consistent with the same molecular-hydrogen Kennicutt-Schmidt law that applies within the optical disk. However, the most significant outliers may reflect evolutionary effects in our H$\alpha$- or FUV- selected regions (Figure~\ref{schmidt}). \\

\noindent 2. We found no correlation between the age of the XUV disk star forming regions based on the H$\alpha$-to-FUV flux ratio ($1-7 \, {\rm Myr}$) and the CO detection rate (Figure~\ref{age}). \\

Deeper CO data for more XUV disk star forming regions would facilitate a number of investigations. We could determine the range of molecular gas SFEs in XUV disk star forming regions, test our hypothesis that molecular gas should be present in young regions like those we studied ($< 7 \, {\rm Myr}$), and determine which conditions control the CO detection rate. Finally, we may be able to use the presence or absence of molecular gas as a coarse age probe to break the degeneracy between age and stochastic sampling of the IMF as the explanation for the low H$\alpha$-to-FUV flux ratios in XUV disks.

\section*{Acknowledgements}

We thank Armando Gil de Paz for his advice throughout the project, Andreas Schruba for sharing his CO(2--1) radial profile, Annette Ferguson for sharing her H$\alpha$ image of NGC~6946, Jonathan Braine for sharing his CO line intensities for NGC~4414, and the referee for comments that improved this work. UL has been supported by the research projects AYA2011-24728 from the Spanish Ministerio de Ciencia y Educaci\'on and the Junta de Andaluc\'\i a (Spain) grants FQM108. This work is based on observations carried out with the IRAM 30m Telescope. IRAM is supported by INSU/CNRS (France), MPG (Germany) and IGN (Spain). This work made use of HERACLES, ``The HERA CO-Line Extragalactic Survey'' \citep{leroy09}. Some of the data presented in this paper were obtained from the Mikulski Archive for Space Telescopes (MAST). STScI is operated by the Association of Universities for Research in Astronomy, Inc., under NASA contract NAS5-26555. Support for MAST for non-HST data is provided by the NASA Office of Space Science via grant NNX09AF08G and by other grants and contracts. This research has made use of the NASA/IPAC Extragalactic Database (NED) which is operated by the Jet Propulsion Laboratory, California Institute of Technology, under contract with the National Aeronautics and Space Administration.

\begin{landscape}
\begin{deluxetable}{lcccccccccc}
\setlength{\tabcolsep}{5.25pt}
\tablewidth{0pt}
\tabletypesize{\scriptsize}
\tablecaption{Measured and Literature Properties Related to Molecular-Hydrogen Kennicutt-Schmidt Law}
\tablehead{
\colhead{Region} &
\colhead{RA} &
\colhead{DEC} &
\colhead{$r/r_{25}$} &
\colhead{$A_{V}$} &
\colhead{$f_{[NII]}$} &
\colhead{$I_{\rm CO(1-0)}$} &
\colhead{$I_{\rm CO(2-1)}$} &
\colhead{$f_{\rm H\alpha}^{21\arcsec}$} &
\colhead{$f_{\lambda,\rm{FUV}}^{21\arcsec}$} &
\colhead{$I_{24 \, \mu {\rm m}}^{21\arcsec}$} \\
\colhead{} &
\colhead{(hh:mm:ss.s)} &
\colhead{(dd:mm:ss)} &
\colhead{} &
\colhead{} &
\colhead{} &
\colhead{(${\rm K \, km\, s^{-1}}$)} &
\colhead{(${\rm K \, km\, s^{-1}}$)} &
\colhead{($10^{-15} \, {\rm erg \, s^{-1} \, cm^{-2}}$)} &
\colhead{($10^{-16} \, {\rm erg \, s^{-1} \, cm^{-2} \, \AA^{-1}}$)} &
\colhead{(${\rm MJy \, sr^{-1}}$)} \\
\colhead{(1)} &
\colhead{(2)} &
\colhead{(3)} &
\colhead{(4)} &
\colhead{(5)} &
\colhead{(6)} &
\colhead{(7)} &
\colhead{(8)} &
\colhead{(9)} &
\colhead{(10)} &
\colhead{(11)} 
}
\startdata
\hline
\multicolumn{11}{c}{NGC 4625} \\
\hline
R2  &    12:42:05.136 &	+41:16:12.14 &	3.4  &	0.048 &  0.894 & $ <0.05 $        & $	<0.04 $          & $ >1.5	  $ & $ 5.9 \pm 0.6 $ & $ <0.018 $         \\
R3  &    12:42:04.025 &	+41:16:35.98 &	3.1  &	0.048 &  0.878 &    ...	           &	...              & $ 1.0  \pm 0.2 $ & $ 3.0 \pm 0.3 $ & $ <0.018 $         \\  
R4  &    12:42:00.322 &	+41:16:39.80 &	2.1  &	0.049 &  0.859 &    ...	           & $	<0.6 $	         & $ 0.95 \pm 0.19$ & $ 2.4 \pm 0.3 $ & $ <0.018 $         \\  
R5  &    12:41:59.150 &	+41:15:21.11 &	2.4  &	0.050 &  0.829 &    ...	           & $	<0.6 $	         & $ 1.8  \pm 0.4 $ & $ 3.0 \pm 0.3 $ & $ <0.018 $         \\
R6  &    12:41:57.029 &	+41:15:53.65 &	1.4  &	0.050 &  0.829 &    ...	           & $	<0.6 $	         & $ 5.3  \pm 1.1 $ & $ 9.4 \pm 1.0 $ & $ 0.072 \pm 0.007 $ \\  
R7  &    12:41:56.234 &	+41:15:30.90 &	1.6  &	0.050 &  0.902 &    ...	           & $	<0.6 $	         & $ 2.6  \pm 0.5 $ & $ 2.6 \pm 0.3 $ & $ <0.018 $         \\  
R8  &    12:41:54.310 &	+41:17:26.82 &	1.5  &	0.049 &  0.880 &    ...	           & $	<0.6 $	         & $ 2.3  \pm 0.5 $ & $ 2.8 \pm 0.3 $ & $ <0.018 $         \\  
R9  &    12:41:51.538 &	+41:15:17.52 &	1.7  &	0.050 &  0.842 &    ...	           & $	<0.6 $           & $ 1.5  \pm 0.3 $ & $ 2.4 \pm 0.3 $ & $ <0.018 $         \\
R10 &    12:41:48.677 &	+41:17:46.81 &	2.3  &	0.049 &  0.936 &    ...	           & $	<0.6 $	         & $ 4.0  \pm 0.8 $ & $ 4.6 \pm 0.5 $ & $ <0.018 $         \\ 
R11 &    12:41:45.151 &	+41:16:00.29 &	2.3  &	0.051 &  0.919 &    ...	           & $	<0.6 $	         & $ 4.2  \pm 0.8 $ & $ 2.5 \pm 0.3 $ & $ <0.018 $         \\ 
R12 &    12:41:42.977 &	+41:16:37.95 &	1.9  &	0.051 &  0.904 &    ...	           & $	<0.6 $	         & $ 4.4  \pm 0.9 $ & $ 5.5 \pm 0.6 $ & $ <0.018 $         \\ 
\hline
\multicolumn{11}{c}{NGC 6946} \\
\hline
P7  &	20:34:26.6    &      +60:12:47.9   &	1.0  &	0.951 &	0.903  & $ 0.21 \pm 0.03 $ & $	0.22 \pm 0.05 $	 & $ 11   \pm  2    $ & $ <5	        $ & $ 0.13  \pm 0.03  $ \\
P2  &	20:34:29.0    &      +60:13:50.5   &	1.1  &	0.956 &	0.903  & $ 0.14 \pm 0.02 $ & $	0.14 \pm 0.03 $	 & $ 150  \pm  30 $ & $ 70 \pm 8  $ & $ 0.34  \pm 0.07  $ \\
P8  &	20:34:54.3    &      +60:15:40.5   &	1.3  &	0.979 &	0.903  & $ 0.15 \pm 0.02 $ & $	<0.3                $	 & $ 22   \pm  4    $ & $ <6	        $ & $ 0.096 \pm 0.019 $ \\
P3  &	20:33:48.4    &      +60:08:32.6   &	1.4  &	0.876 &	0.903  & $ 0.11 \pm 0.02 $ & $	<0.13	       $	 & $ 23   \pm  5    $ & $ 31 \pm 3  $ & $ <0.08	          $ \\ 
P5  &	20:34:51.4    &      +60:16:24.9   &	1.5  &	0.987 &	0.903  & $ <0.13              $ & $	<0.5                $       & $ 3.7  \pm 0.7  $ & $ <6	        $ & $ <0.08	          $ \\ 
P4  &	20:34:51.4    &      +60:16:44.9   &	1.5  &	0.991 &	0.903  & $ <0.14              $ & $	<0.4                $	 & $ <1.6              $ & $ <6	        $ & $ <0.08	          $ \\
\hline
\multicolumn{11}{c}{M33} \\
\hline
M33\_18.1  &	 01:34:13.31 &      +31:10:32.0  &	1.1  &	0.129 &	  ...      & $ 0.52 \pm 0.06 $ & $ 0.40 \pm 0.05 $	 &  ... &  $ 72   \pm    8   $  & $  <0.03  $ \\
M33\_18.2  &	 01:34:11.94 &      +31:10:32.0  &	1.1  &	0.128 &	  ...      & $ 0.26 \pm 0.06 $ & ...	                         &  ... &  $ 6.2  \pm    0.7$  & $  <0.03  $ \\
M33\_18.3  &	 01:34:13.31 &      +31:10:21.5  &	1.1  &	0.128 &	  ...      & $ 0.24 \pm 0.03 $ & ...                      	 &  ... &  $ 42   \pm    5    $ & $  <0.03  $ \\
M33\_18.4  &	 01:34:12.61 &      +31:10:32.0  &	1.1  &	0.128 &	  ...      & $ 0.61 \pm 0.02 $ & $ 1.22 \pm 0.03 $	 &  ... &  $ 43   \pm    5    $ & $  <0.03  $ \\ 
M33\_20.1  &	 01:34:47.79 &      +31:08:23.8  &	1.0  &	0.135 &	  ...      & $ 0.26 \pm 0.03 $ & ...                              &  ...  &  $ 27   \pm    3    $ & $  <0.03  $ \\ 
M33\_20.2  &	 01:34:49.79 &      +31:08:23.8  &	1.0  &	0.135 &	  ...      & $ 0.33 \pm 0.04 $ & ...                      	 &  ... &   $10.1 \pm   1.1  $ & $  <0.03  $ \\
\hline
\multicolumn{11}{c}{NGC 4414} \\
\hline
R1  &	 12:26:30.51 &       +31:11:06.9 &	1.4  &	0.054 &	  ...      & $ 0.11 \pm 0.02 $ & ...	 & ...  &   $ 1.8   \pm  0.2  $ & $ <0.5 $ \\
R2  &	 12:26:29.17 &       +31:11:30.0 &	1.1  &	0.054 &	  ...      & $ 0.15 \pm 0.02 $ & ...	 & ...  &   $ 2.2   \pm  0.2  $ & $ <0.5 $ \\
R3  &	 12:26:26.57 &       +31:15:37.7 &	1.4  &	0.053 &	  ...      & $ 0.23 \pm 0.03 $ & ...	 & ...  &   $ 3.1    \pm 0.3  $ & $ <0.5 $ \\
\enddata

\tablecomments{Column 1: Region name, as originally presented in \citet{gildepaz07} for NGC~4625 and \citet{braine07} for NGC~6946. For M33, the names are from \citet{braine10} and we added the numbers after the decimal point. R1, R2, and R3 in NGC~4414 correspond to offset positions (51, -138), (31, -115), and (-8, 133), respectively, from \citet{braine04}. We did not include offset position (-68, 162) because it falls outside the FUV image. Column 2 and 3: RA and DEC (J2000.0) of regions.  Column 4: Galactocentric radius, normalized to $r_{25}$, which is $41.4\arcsec$ for NGC~4625 \citep{schruba11}, $345.0\arcsec$ for NGC~6946 \citep{braine07}, $1848\arcsec$ for M33 \citep{paturel03}, and $108.9\arcsec$ for NGC~4414 \citep{devaucouleurs91}. The NGC~4625 galactocentric radii are from \citet{gildepaz07}. We computed the NGC~6946, M33, and NGC~4414 values from the region coordinates in Columns 2 and 3, the position angles and inclinations from \citet{devaucouleurs91}, and the galaxy centers from \citet{munoz09} or the NASA/IPAC Extragalactic Database (NED). Note that use of $r_{25}$ from \citet{devaucouleurs91} puts the M33 regions at $r<r_{25}$. Column 5: Extinction in the V-band due to the Galaxy \citep{schlafly11}. Column 6: Multiplicative factor applied to obtain the H$\alpha$ flux from the measured flux, which includes H$\alpha$ and [NII] $\lambda \lambda 6548,6584$ emission. For NGC~4625, the [NII] correction factor is based on the [NII] $\lambda 6584$/H$\alpha$ ratios from \citet{gildepaz07} and we assumed the theoretical value of [NII]$\lambda 6548$/[NII]$\lambda 6584$ of 0.33 \citep{osterbrock89}. For R3, we used the average ratio from the remaining regions. For NGC~6946, we used a constant [NII] correction factor, based on the measured value for P3 from \citet{ferguson98a}. Column 7 and 8: CO(1-0) and CO(2-1) line intensity, within the 21$\arcsec$- and 11$\arcsec$-diameter beams, respectively. All upper limits in the table are 3$\sigma$ upper limits. The upper limit for R2 in NGC~4625 is from this work, assuming a line width of $18.2 \, {\rm km \, s^{-1}}$. We calculated the upper limits for the remaining regions in NGC 4625 from the HERACLES survey integrated intensity map \citep{leroy09}. The values for regions in NGC~6946 and M33 are from \citet{braine07} and \citet{braine10}, respectively, where the uncertainties are based on the rms noise in the spectrum and upper limits were calculated with a line width of $18.2 \, {\rm km \, s^{-1}}$. The NGC~4414 values are from \citet{braine04} and we assumed 15\% uncertainty. Column 9: H$\alpha$ flux within an 21$\arcsec$-diameter aperture, corrected for [NII] emission and Galactic extinction (not internal extinction). Column 10: FUV flux density within a 21$\arcsec$-diameter aperture, corrected for Galactic extinction. Column 11: Average $24 \, {\rm \mu m}$ surface brightness within a 21$\arcsec$-diameter aperture.}

\label{tab:obs_prop}
\end{deluxetable}
\end{landscape}

\begin{deluxetable}{lccccccc}
\setlength{\tabcolsep}{1pt}
\tablewidth{0pt}
\tabletypesize{\scriptsize}
\tablecaption{Measured and Literature Properties Related to Age Estimate}
\tablehead{
\colhead{Region} &
\colhead{Offset} &
\colhead{$E({\rm B-V})$} &
\colhead{$12 + {\rm log(O/H)}$} &
\colhead{$A_{\rm FUV}$} &
\colhead{$f_{\rm H\alpha}^{6\arcsec}$} &
\colhead{$f_{\lambda,\rm{FUV}}^{6\arcsec}$} &
\colhead{$L_{\nu,\rm{FUV}}^{6\arcsec}$} \\
\colhead{} &
\colhead{(arcsec)} &
\colhead{} &
\colhead{} &
\colhead{} &
\colhead{($10^{-15} \, {\rm erg \, s^{-1} \, cm^{-2}}$)} &
\colhead{($10^{-16} \, {\rm erg \, s^{-1} \, cm^{-2} \, \AA^{-1}}$)} &
\colhead{($10^{24} \, {\rm erg \, s^{-1} \, Hz^{-1}}$)} \\
\colhead{(1)} &
\colhead{(2)} &
\colhead{(3)} &
\colhead{(4)} &
\colhead{(5)} &
\colhead{(6)} &
\colhead{(7)} &
\colhead{(8)}
}
\startdata
\hline
\multicolumn{8}{c}{NGC 4625} \\
\hline
R2  &    0.0  &	0.31  & 8.531 &	  1.3	          &	$   3.5  \pm 0.7  $   &	$   5.3  \pm 0.6  $   &	$   4.4  \pm 0.5  $   \\
R3  &    0.0  &	0.23  & ...   &	  1.0	          &	$   1.1  \pm 0.2  $   &	$   1.11 \pm 0.12 $   &	$   0.92 \pm 0.10 $   \\
R4  &    0.0  &	0.54  & ...   &	  2.4	          &	$   3.0  \pm 0.6  $   &	$   7.6  \pm 0.8  $   &	$   6.3  \pm 0.7  $   \\
R5  &    0.0  &	0.13  & 8.123 &	  0.5	          &	$   1.7  \pm 0.3  $   &	$   1.8  \pm 0.2  $   &	$   1.47 \pm 0.16 $   \\
R6  &    0.0  &	0.25  & 8.265 &	  1.1	          &	$   6.3  \pm 1.3  $   &	$   6.7  \pm 0.7  $   &	$   5.6  \pm 0.6  $   \\
R7  &    0.0  &	0.29  & 8.296 &	  1.2	          &	$   3.8  \pm 0.8  $   &	$   1.58 \pm 0.17 $   &	$   1.30 \pm 0.14 $   \\
R8  &    0.0  &	0.45  & ...   &	  2.0	          &	$   2.2  \pm 0.4  $   &	$   2.8  \pm 0.3  $   &	$   2.3  \pm 0.3  $   \\
R9  &    0.0  &	0.30  & 8.400 &	  1.3	          &	$   0.92 \pm 0.18 $   &	$   0.90 \pm 0.10 $   &	$   0.75 \pm 0.08 $   \\
R10 &    0.0  &	0.00  & ...   &	  0.0	          &	$   3.4  \pm 0.7  $   &	$   1.06 \pm 0.12 $   &	$   0.88 \pm 0.10 $   \\
R11 &    0.0  &	0.66  & ...   &	  2.9	          &	$   10   \pm 2    $   &	$   11.4 \pm 1.3  $   &	$   9.5  \pm 1.0  $   \\
R12 &    0.0  &	0.22  & 8.333 &	  0.9	          &	$   3.0  \pm 0.6  $   &	$   3.7  \pm 0.4  $   &	$   3.0  \pm 0.3  $   \\
\hline
\multicolumn{8}{c}{NGC 6946} \\
\hline
P7  &	 3.0  & ...   & ...   &	$ 1.13 \pm 0.16 $ &	$   4.4  \pm 1.3  $   &	$   <4	          $   &	$   <1.4	  $   \\
P2  &	 1.1  &	...   & ...   &	$ 1.0  \pm 0.3  $ &	$   110  \pm 30   $   &	$   110  \pm 40   $   &	$   34   \pm 12   $   \\
P8  &	 3.6  &	...   & ...   &	$ 0.7  \pm 0.3  $ &	$   17   \pm 5    $   &	$   <3	          $   &	$   <1.0	  $   \\
P3  &	 1.4  & ...   & 8.1   &	$ 0.6  \pm 0.5  $ &	$   16   \pm 5    $   &	$   23   \pm 8    $   &	$   7    \pm 3    $   \\
P5  &	 0.8  & ...   & ...   &	$ 0.7  \pm 0.5  $ &	$   1.7  \pm 0.5  $   &	$   <3	          $   &	$   <1.0	  $   \\
P4  &	 0.0  &	...   & ...   &	$ 0.8  \pm 0.9  $ &	$   <0.7	  $   &	$   <4	          $   &	$   <1.2	  $   \\
\enddata

\tablecomments{Column 1: Region name, as originally presented in
  \citet{gildepaz07} for NGC~4625 and \citet{braine07} for
  NGC~6946. Column 2: Offset between the aperture center for the
  H$\alpha$ and FUV flux measurements within the $6\arcsec$-diameter
  aperture and the coordinates from \citet[][NGC~4625]{gildepaz07} or
  \citet[][NGC~6946]{braine07}. Column 3: Color excess for NGC~4625
  from \citet{gildepaz07}, based on the Balmer line decrement and
  including reddening due to both Galactic and internal dust. Column
  4: $12+{\rm log(O/H)}$ from \citet{gildepaz07} for NGC~4625
  and \citet{ferguson98a} for NGC~6946. All values are based on the 
  R23 calibration using the \citet{mcgaugh91} method. Column
  5: FUV extinction, due to internal extinction only. For NGC~4625, we
  derived this from $A_{V}$ from Column~5 of Table~\ref{tab:obs_prop},
  $E(B-V)$ from Column 3, and the \citet{calzetti01} relation $A_{\rm
  FUV} = 1.8 \, A_{\rm H\alpha}$. For NGC~6946, we used the
  \citet{munoz09} $A_{\rm FUV}$ radial profile based on the
  relationship between internal FUV extinction and the TIR-to-FUV
  luminosity ratio. Column 6: H$\alpha$ flux within a
  6$\arcsec$-diameter aperture, corrected for [NII] emission and
  Galactic and internal extinction. For NGC~4625, we corrected for
  Galactic and internal extinction with the $E(B-V)$ from Column
  3. For NGC~6946, we corrected for Galactic and internal extinction
  with $A_{V}$ from Column~5 of Table~\ref{tab:obs_prop}, $A_{\rm
  FUV}$ from Column~5, and the \citet{calzetti01} relation $A_{\rm
  FUV} = 1.8 \, A_{\rm H\alpha}$. All upper limits in the table are
  $3\sigma$ upper limits. Column 7: FUV flux density within a
  6$\arcsec$-diameter aperture, corrected for Galactic and internal
  extinction with $A_{V}$ from Column~5 of Table~\ref{tab:obs_prop}
  and $A_{\rm FUV}$ from Column~5. Column 8: FUV luminosity, assuming
  a distance of $9.5 \, {\rm Mpc}$ for NGC~4625 and $5.9 \, {\rm Mpc}$
  for NGC~6946 \citep{schruba11}.}

\label{tab:obs_prop2}
\end{deluxetable}


\begin{thebibliography}{99}

\bibitem[Alberts et al.(2011)]{alberts11} Alberts, S., Calzetti, D.,
Dong, H., et al.\ 2011, ApJ, 731, 28

\bibitem[Albrecht et al.(2004)]{albrecht04} Albrecht, M., Chini, R.,
  Kr{\"u}gel, E., M{\"u}ller, S.~A.~H., \& Lemke, R.\ 2004, A\&A, 414,
  141

\bibitem[Barnes et al.(2012)]{barnes12} Barnes, K.~L., van Zee, L.,
C{\^o}t{\'e}, S., \& Schade, D.\ 2012, ApJ, 757, 64

\bibitem[Bigiel et al.(2008)]{bigiel08} Bigiel, F., Leroy, A., Walter,
F., Brinks, E., de Blok, W.~J.~G., Madore, B., \& Thornley, M.~D.\
2008, AJ, 136, 2846

\bibitem[Bigiel et al.(2010)]{bigiel10} Bigiel, F., Leroy, A., Walter,
F., et al.\ 2010, AJ, 140, 1194

\bibitem[Blanc et al.(2009)]{blanc09} Blanc, G.~A., Heiderman, 
A., Gebhardt, K., Evans, N.~J., II, \& Adams, J.\ 2009, ApJ, 704, 842 

\bibitem[Blitz \& Rosolowsky(2006)]{blitz06} Blitz, L., \& Rosolowsky, E.\ 2006, ApJ, 650, 933 

\bibitem[Boissier et al.(2005)]{boissier05}  Boissier, S., Gil de Paz,
A., Madore, B.~F., et al.\ 2005, ApJ, 619, L83

\bibitem[Boissier et al.(2007)]{boissier07} Boissier, S., Gil de Paz,
A., Boselli, A., et al.\ 2007, ApJS, 173, 524

\bibitem[Bolatto et al.(2013)]{bolatto13} Bolatto, A.~D., Wolfire, M., \& Leroy, A.~K.\ 2013, ARA\&A, 51, 207 

\bibitem[Bonanos et al.(2006)]{bonanos06} Bonanos, A.~Z., Stanek, 
K.~Z., Kudritzki, R.~P., et al.\ 2006, ApJ, 652, 313 

\bibitem[Braine \& Herpin(2004)]{braine04} Braine, J., \& Herpin, F.\ 2004, Nature, 432, 369 

\bibitem[Braine et al.(2007)]{braine07} Braine, J., Ferguson,
A.~M.~N., Bertoldi, F., \& Wilson, C.~D.\ 2007, ApJ, 669, L73

\bibitem[Braine et al.(2010)]{braine10} Braine, J., Gratier, P., Kramer, C., et al.\ 2010, A\&A, 520, A107 

\bibitem[Bush \& Wilcots(2004)]{bush04} Bush, S.~J., \& Wilcots,
E.~M.\ 2004, AJ, 128, 2789

\bibitem[Calzetti(2001)]{calzetti01} Calzetti, D.\ 2001, PASP, 113,
1449

\bibitem[Cardelli et al.(1989)]{cardelli89} Cardelli, J.~A., Clayton,
G.~C., \& Mathis, J.~S.\ 1989, ApJ, 345, 245

\bibitem[Carilli \& Walter(2013)]{carilli13} Carilli, C.~L., \& Walter, F.\ 2013, ARA\&A, 51, 105 

\bibitem[Daddi et al.(2010)]{daddi10} Daddi, E., Elbaz, D., Walter, F., et al.\ 2010, ApJ, 714, L118 

\bibitem[Dale et al.(2009)]{dale09} Dale, D.~A., Cohen, S.~A., Johnson, L.~C., et al.\ 2009, ApJ, 703, 517 

\bibitem[Dessauges-Zavadsky et 
al.(2014)]{dessauges14} Dessauges-Zavadsky, M., Verdugo, C., Combes, F., \& Pfenniger, D.\ 2014, A\&A, 566, A147 

\bibitem[de Vaucouleurs et al.(1991)]{devaucouleurs91}de  Vaucouleurs,  G.,  de  Vaucouleurs,  A.,  Corwin,  H.  G.,  et  al.  1991,  Third Reference Catalogue of Bright Galaxies (New York: Springer)

\bibitem[Dobbs \& Pringle(2013)]{dobbs13} Dobbs, C.~L., \& Pringle, J.~E.\ 2013, MNRAS, 432, 653 

\bibitem[Dong et al.(2008)]{dong08} Dong, H., Calzetti, D., Regan, M.,
et al.\ 2008, AJ, 136, 479

\bibitem[Dopita et al.(2006)]{dopita06} Dopita, M.~A., Fischera, J.,
Sutherland, R.~S., et al.\ 2006, ApJS, 167, 177

\bibitem[Elmegreen \& Hunter(2006)]{elmegreen06} Elmegreen, B.~G., \&
Hunter, D.~A.\ 2006, ApJ, 636, 712

\bibitem[Ferguson et al.(1998a)]{ferguson98a} Ferguson, A.~M.~N.,
Gallagher, J.~S., \& Wyse, R.~F.~G.\ 1998a, AJ, 116, 673

\bibitem[Ferguson et al.(1998b)]{ferguson98b} Ferguson, A.~M.~N., 
Wyse, R.~F.~G., Gallagher, J.~S., \& Hunter, D.~A.\ 1998b, ApJ, 506, L19

\bibitem[Freedman et al.(2001)]{freedman01} Freedman, W.~L., 
Madore, B.~F., Gibson, B.~K., et al.\ 2001, ApJ, 553, 47 

\bibitem[Fukui \& Kawamura(2010)]{fukui10} Fukui, Y., \& Kawamura, A.\
2010, ARA\&A, 48, 547

\bibitem[Genzel et al.(2010)]{genzel10} Genzel, R., Tacconi, 
L.~J., Gracia-Carpio, J., et al.\ 2010, MNRAS, 407, 2091 

\bibitem[Gil de Paz et al.(2005)]{gildepaz05} Gil de Paz, A., Madore,
B.~F., Boissier, S., et al.\ 2005, ApJ, 627, L29

\bibitem[Gil de Paz et al.(2007a)]{gildepaz07} Gil de Paz, A., Madore,
B.~F., Boissier, S., et al.\ 2007a, ApJ, 661, 115

\bibitem[Gil de Paz et al.(2007b)]{gildepaz07b} Gil de Paz, A.,
Boissier, S., Madore, B.~F., et al.\ 2007b, ApJS, 173, 185

\bibitem[Glover \& Mac Low(2011)]{glover11} Glover, S.~C.~O., \& Mac Low, M.-M.\ 2011, MNRAS, 412, 337 

\bibitem[Goddard et al.(2010)]{goddard10} Goddard, Q.~E., Kennicutt,
R.~C., \& Ryan-Weber, E.~V.\ 2010, MNRAS, 405, 2791

\bibitem[Goddard et al.(2011)]{goddard11} Goddard, Q.~E., Bresolin,
F., Kennicutt, R.~C., Ryan-Weber, E.~V., \& Rosales-Ortega, F.~F.\
2011, MNRAS, 412, 1246

\bibitem[Gogarten et al.(2009)]{gogarten09} Gogarten, S.~M.,
Dalcanton, J.~J., Williams, B.~F., et al.\ 2009, ApJ, 691, 115

\bibitem[Hartman et al.(2006)]{hartman06} Hartman, J.~D., 
Bersier, D., Stanek, K.~Z., et al.\ 2006, MNRAS, 371, 1405 

\bibitem[Holwerda et al.(2012)]{holwerda12} Holwerda, B.~W., 
Pirzkal, N., \& Heiner, J.~S.\ 2012, MNRAS, 427, 3159 

\bibitem[Hoversten \& Glazebrook(2008)]{hoversten08} Hoversten, E.~A., \& Glazebrook, K.\ 2008, ApJ, 675, 163 

\bibitem[Hunter et al.(2013)]{hunter13} Hunter, D.~A., 
Elmegreen, B.~G., Rubin, V.~C., et al.\ 2013, AJ, 146, 92 

\bibitem[Kawamura et al.(2009)]{kawamura09} Kawamura, A., Mizuno, 
Y., Minamidani, T., et al.\ 2009, ApJS, 184, 1 

\bibitem[Kennicutt(1998)]{kennicutt98} Kennicutt, R.~C., Jr.\ 1998, 
ApJ, 498, 541 

\bibitem[Kennicutt et al.(2003)]{kennicutt03} Kennicutt, R.~C., Jr.,
Armus, L., Bendo, G., et al.\ 2003, PASP, 115, 928

\bibitem[Kennicutt et al.(2008)]{kennicutt08} Kennicutt, R.~C., Jr.,
Lee, J.~C., Funes, S.~J., Jos{\'e} G., Sakai, S., \& Akiyama, S.\
2008, ApJS, 178, 247

\bibitem[Kennicutt \& Evans(2012)]{kennicutt12} Kennicutt, R.~C., \&
Evans, N.~J.\ 2012, ARA\&A, 50, 531

\bibitem[Koda et al.(2012)]{koda12} Koda, J., Yagi, M., Boissier, S.,
et al.\ 2012, ApJ, 749, 20

\bibitem[Kruijssen \& Longmore(2014)]{kruijssen14} Kruijssen, J.~M.~D., \& Longmore, S.~N.\ 2014, MNRAS, 439, 3239 

\bibitem[Lee et al.(2009)]{lee09} Lee, J.~C., Gil de Paz, A.,
Tremonti, C., et al.\ 2009, ApJ, 706, 599

\bibitem[Leitherer et al.(1999)]{leitherer99} Leitherer, C., 
Schaerer, D., Goldader, J.~D., et al.\ 1999, ApJS, 123, 3 

\bibitem[Lemonias et al.(2011)]{lemonias11} Lemonias, J.~J.,
Schiminovich, D., Thilker, D., et al.\ 2011, ApJ, 733, 74

\bibitem[Leroy et al.(2008)]{leroy08} Leroy, A.~K., Walter, F., 
Brinks, E., et al.\ 2008, AJ, 136, 2782 

\bibitem[Leroy et al.(2009)]{leroy09} Leroy, A.~K., Walter, F.,
Bigiel, F., et al.\ 2009, AJ, 137, 4670

\bibitem[Leroy et al.(2011)]{leroy11} Leroy, A.~K., Bolatto, 
A., Gordon, K., et al.\ 2011, ApJ, 737, 12 

\bibitem[Leroy et al.(2012)]{leroy12} Leroy, A.~K., Bigiel, F., de
Blok, W.~J.~G., et al.\ 2012, AJ, 144, 3

\bibitem[Leroy et al.(2013)]{leroy13} Leroy, A.~K., Walter, F., 
Sandstrom, K., et al.\ 2013, AJ, 146, 19 

\bibitem[Martin \& Kennicutt(2001)]{martin01} Martin, C.~L., \&
Kennicutt, R.~C., Jr.\ 2001, ApJ, 555, 301

\bibitem[McGaugh(1991)]{mcgaugh91} McGaugh, S.~S.\ 1991, ApJ, 
380, 140

\bibitem[McKee \& Ostriker(2007)]{mckee07} McKee, C.~F., \& Ostriker, E.~C.\ 2007, ARA\&A, 45, 565 

\bibitem[Meurer et al.(2009)]{meurer09} Meurer, G.~R., Wong, O.~I., Kim, J.~H., et al.\ 2009, ApJ, 695, 765 

\bibitem[Moffett et al.(2012)]{moffett12} Moffett, A.~J., Kannappan,
S.~J., Baker, A.~J., \& Laine, S.\ 2012, ApJ, 745, 34

\bibitem[Morrissey et al.(2007)]{morrissey07} Morrissey, P., Conrow,
T., Barlow, T.~A., et al.\ 2007, ApJS, 173, 682

\bibitem[Mu{\~n}oz-Mateos et al.(2009)]{munoz09} Mu{\~n}oz-Mateos,
J.~C., Gil de Paz, A., Boissier, S., et al.\ 2009, ApJ, 701, 1965

\bibitem[Narayanan et al.(2012)]{narayanan12} Narayanan, D., 
Krumholz, M.~R., Ostriker, E.~C., \& Hernquist, L.\ 2012, MNRAS, 421, 3127 

\bibitem[O'Donnell(1994)]{odonnell94} O'Donnell, J.~E.\ 1994, ApJ,
422, 158

\bibitem[Osterbrock(1989)]{osterbrock89} Osterbrock, D.~E.\ 1989, 
Astrophysics of Gaseous Nebulae and Active Galactic
Nuclei (Mill Valley, CA: University Science Books)

\bibitem[Paturel et al.(2003)]{paturel03} Paturel, G., Petit, C., Prugniel, P., et al.\ 2003, A\&A, 412, 45 

\bibitem[Rahman et al.(2012)]{rahman12} Rahman, N., Bolatto, 
A.~D., Xue, R., et al.\ 2012, ApJ, 745, 183 

\bibitem[Sandstrom et al.(2013)]{sandstrom13} Sandstrom, K.~M., 
Leroy, A.~K., Walter, F., et al.\ 2013, ApJ, 777, 5 

\bibitem[Schlafly \& Finkbeiner(2011)]{schlafly11} Schlafly, E.~F., \&
Finkbeiner, D.~P.\ 2011, ApJ, 737, 103

\bibitem[Schlegel et al.(1998)]{schlegel98} Schlegel, D.~J.,
Finkbeiner, D.~P., \& Davis, M.\ 1998, ApJ, 500, 525

\bibitem[Schmidt(1959)]{schmidt59} Schmidt, M.\ 1959, ApJ, 129, 243

\bibitem[Schruba et al.(2010)]{schruba10} Schruba, A., Leroy, 
A.~K., Walter, F., Sandstrom, K., \& Rosolowsky, E.\ 2010, ApJ, 722, 1699 

\bibitem[Schruba et al.(2011)]{schruba11} Schruba, A., et al.\ 2011,
AJ, 142, 37

\bibitem[Snell et al.(2002)]{snell12} Snell, R.~L., Carpenter, J.~M.,
\& Heyer, M.~H.\ 2002, ApJ, 578, 229

\bibitem[Thilker et al.(2005)]{thilker05} Thilker, D.~A., Bianchi, L.,
Boissier, S., et al.\ 2005, ApJ, 619, L79

\bibitem[Thilker et al.(2007)]{thilker07} Thilker, D.~A., Bianchi, L.,
Meurer, G., et al.\ 2007, ApJS, 173, 538

\bibitem[Tremonti et al.(2004)]{tremonti04} Tremonti, C.~A., 
Heckman, T.~M., Kauffmann, G., et al.\ 2004, ApJ, 613, 898 

\bibitem[Whitmore et al.(2011)]{whitmore11} Whitmore, B.~C., Chandar,
R., Kim, H., et al.\ 2011, ApJ, 729, 78

\bibitem[Wolfire et al.(2010)]{wolfire10} Wolfire, M.~G., 
Hollenbach, D., \& McKee, C.~F.\ 2010, ApJ, 716, 1191 

\bibitem[Wong \& Blitz(2002)]{wong02} Wong, T., \& Blitz,
L.\ 2002, ApJ, 569, 157

\end{thebibliography}
\end{document}